\documentclass[aps,prc,twocolumn,groupedaddress,floatfix]{revtex4-1}
\usepackage{bm}
\usepackage{color}
\usepackage{graphicx}
\usepackage{amssymb,amsmath}
\allowdisplaybreaks[2]
\usepackage{multirow}
\usepackage{subfigure}
\newcommand{\Het}{{}^3{\rm He}}

\newcommand{\Ht}{{}^3{\rm H}}
\newcommand{\Li}{{}^6{\rm Li}}
\newcommand{\xx}{\boldsymbol{\xi}}
\newcommand{\hxx}{\hat{\xi}}
\newcommand{\ph}{\varphi}
\newcommand{\br}{\boldsymbol{r}}
\newcommand{\brr}{\boldsymbol{R}}
\newcommand{\al}{\alpha}
\newcommand{\bra}{\langle}
\newcommand{\ket}{\rangle}
\newcommand{\phz}{\phantom{0}}
\newcommand{\kb}{{\overline{K}}}
\newcommand{\lb}{{\overline{L}}}
\newcommand{\sbb}{{\overline{S}}}
\newcommand{\tb}{{\overline{T}}}
\newcommand{\bb}{{\overline{\alpha}}}
\newcommand{\lbb}{{\overline{l}}}

\begin{document}


\title{Calculation of the $\Li$ ground state within the hyperspherical
  harmonic basis}


\author{A.\ Gnech$^{\,{\rm a,b}}$, M.\ Viviani$^{\,{\rm b}}$,
  L.\ E.\ Marcucci$^{\, {\rm c,b}}$}
\affiliation{
$^{\rm a}$\mbox{Gran Sasso Science Institute, 67100 L'Aquila, Italy}\\
$^{\rm b}$\mbox{INFN-Pisa, 56127 Pisa, Italy}\\
$^{\rm c}$\mbox{Universit\'a di Pisa, 56127 Pisa, Italy} 
}


\date{\today}

\begin{abstract}
  We have studied the solution of the six-nucleon bound state
  problem 
  using the hyperspherical harmonic (HH) approach. For this study we have
  considered only two-body nuclear forces. In particular we have used
  a chiral nuclear potential evolved with the
  similarity renormalization group unitary transformation.
  A restricted basis has been selected  by performing a careful
  analysis of the convergence of different HH classes.
  Finally, the binding energy and other properties of $\Li$
  ground state are calculated and compared with the results obtained by
  other techniques.
  Then, we present a calculation of matrix elements relevant
  for direct dark matter search involving $\Li$.
  The results obtained demonstrate the feasibility of using the HH method to perform
  calculation beyond $A=4$.
\end{abstract}


\maketitle
\section{Introduction}
The {\it ab-initio} description of $A=6$ nuclear systems, starting from
realistic
nucleon-nucleon ($NN$) interaction,
requires very large efforts. Indeed, if there are
many very well established methods which are able to solve
the Schr\"odinger equation up to $A=4$ with high precision,
very few {\it ab-initio} techniques can be successfully applied
to the $A=6$ problem.

The methods devised to tackle the problem of the solution
of the non-relativistic Schr\"odinger equation
\begin{equation}\label{eq:eq1}
  H\Psi=E\Psi\,,
\end{equation}
where $H$ is the six-body Hamiltonian, are very different and they usually
are not able to deal with the same potentials.
In the Green's Function Monte Carlo (GFMC) method (see Ref.~\cite{Carlson2015}
and references therein), a trial wave function
is evolved using a stochastic procedure towards the exact ground state
wave function. In this approach only local nuclear interactions
can be used. However, up to now it is the only technique that can reach
a reasonable convergence beyond $A=4$ using realistic interactions
~\cite{Piarulli2018}.
The development of unitary transformation methods,
such as the similarity renormalization group (SRG) method
~\cite{Bogner2007,Jurgenson2009},
and the increasing
power of the computational resources have opened the possibility to
successfully apply variational approaches to system with $A>4$.
The unitary transformation methods permit to soften the short-range
repulsion of the nuclear interactions,
reducing  the dimension of the basis needed to
reach convergence.
The No-Core-Shell-Model (NCSM)  method~\cite{Barrett2013}
performs the calculation by expanding
the wave function on the Harmonic Oscillator (HO) basis and reducing the
problem to an eigenvalue problem. The use of evolved nuclear potentials
does not allow a direct comparison of the NCSM and GFMC results,
because of the emergence in the SRG procedure of many-body
forces that are not usually fully included. 
The effective interaction Hyperspherical Harmonics (EIHH) method~\cite{Barnea2000}
adopts another type of unitary transformation in which the full Hamiltonian of
Eq.~(\ref{eq:eq1}) is replaced with the transformed one, which is easier
to solve, and the calculated eigenvalues quickly converge to the
exact ones.  The results for $A=6$ systems
have been obtained with central potentials~\cite{Barnea2000}, and more recently
with non-local $NN$ interactions~\cite{Barnea2010}.
Other explorative studies in the $A=6$ sector 
using the hyperspherical harmonic (HH)
approach were performed in Refs.~\cite{Vaintraub2009} and~\cite{Gattobigio2011}.   
In Ref.~\cite{Vaintraub2009}, the HH are symmetrized by using the
Casimir operators, and the calculation are performed using the
JISP16~\cite{Shirokov2007}
non-local interaction.
In Ref.~\cite{Gattobigio2011}, the non-symmetrized hyperspherical
harmonics (NSHH) are used with the Volkov~\cite{Volkov1965} potential.
Other results for $\Li$ were obtained by using the stochastic variational
method (SVM)~\cite{Varga1995}, where only central potentials
without the full complication of the modern chiral interactions were used.
Also the solution of $A=5,6$ with
the Faddeev-Yakoubovsky (FY) equation is vigorously
explored~\cite{Lazauskas2019a,Lazauskas2019b}.

In the present work we address the problem of calculating the $\Li$ properties,
using a nuclear Hamiltonian containing only two-body chiral
interactions evolved with the SRG unitary transformation, within the HH approach.
Our goal is to reach reliable convergences 
with these potentials and perform solid extrapolations of the nuclear
observables. The motivation is twofold.
First, we would like to show that the HH approach, as developed by the Pisa group
~\cite{Kievsky2008,Marcucci2019},
can be successfully applied for nuclei with $A>4$. Therefore, this work
can be considered as the first step which will permit to study
a variety of phenomena in $A=6$ sector with the HH method.
Second, we would like to perform comparison with the results obtained by the
NCSM method. In this sense, this work would
represent an important independent
benchmark which validates in $A=6$ sector both the approaches.

The main problem of using the HH expansion is related to the
slow convergence of the basis. In the present case, this problem
is enhanced by the fact that the nucleus of $\Li$ is strongly clustered
in an $\alpha$-particle and a deuteron ($d$). Indeed, the HH basis, being
very compact, has some difficulties in describing clustered
structures. This requires to include a large number of
antisymmetric spin-isospin-HH states
to reach a good convergence even with very soft potential
as in the case of the SRG evolved ones.
Indeed, the expansion in HH states, which is controlled by the
grandangular quantum number $K$ (see below), involves
more than $10^4$ states already for
$K=6$ and they increase exponentially as $K$ grows. It is clear
that a brute force approach is not possible even with sophisticated
computational facilities. The idea is to select a suitable
subsets of states as proposed in Refs.~\cite{Efros1972,Efros1979,Demin1977,Fabre1983,Viviani2005}. Here,
we use an approach very similar to the one used in Ref.~\cite{Viviani2005}
for the $\alpha$ particle, 
where the HH states were selected in terms of the sum of the particles angular
momentum  and the number of correlated particles.
This classification permits to define subsets of states
(hereafter named ``classes'')
having similar properties, so as to optimize
the expansion basis. In the following,
we will carefully analyze the convergence of each class,
so as to have an accurate calculation of the $\Li$ ground state
properties.

The successful application of the HH method is permitted by
the use of the coefficients for the transformation of the
HH functions from a generic permutation of the six particles
to a reference one. To obtain these
coefficients, we
used the same approaches of Refs.~\cite{Viviani1998,Dohet2019}
extending it for six-particles.
The use of the coefficients permits to identify and eliminate
the linear dependent states from the basis. The elimination of
these ``spurious states'' is very useful, since the number of 
linear independent states are noticeably smaller than the full
degeneracy of the basis. Moreover, through the coefficients,
the potential matrix elements can be computed easily and with
high accuracy, performing only
one (two) integration for local (non-local) two-body potentials.

The study presented in this article is the direct
extension of the calculation of Ref.~\cite{Viviani2005} performed on
$A=4$ nucleon systems,
where the use of the same approach for systems with $A>4$
was already predicted. 
The natural continuation will be the inclusion of three-body
forces and the attempt of using also ``bare'' chiral interactions.
These two lines of research are currently underway.
Moreover, we expect that it will be
possible to extend the same approach to nuclei
up to $A=8$.

This article is organized as follows. In Section~\ref{sec:HHexpansion}
we briefly introduce the HH formalism for $A=6$. In Section~\ref{sec:basis}
we present the selection of the HH basis. The convergence of the
binding energy and the study of $\Li$ ground state properties are presented
in Section~\ref{sec:lithres}. In Section~\ref{sec:dm} the matrix elements
relevant for direct dark matter search with $\Li$ are studied.
The last section is dedicated to the conclusions
and the perspectives of the present study. In the Appendix
we present technical details on the algorithm.


\section{The HH expansion}\label{sec:HHexpansion}
The reference set of Jacobi vectors for six equal-mass particles, that we
use in this work, is
  \begin{equation}
  \begin{aligned}
    \xx_{1p}&=\sqrt{\frac{5}{3}}\left(\br_n-\frac{\br_m+\br_l+\br_k+\br_j+\br_i}{5}
    \right)\\
    \xx_{2p}&=\sqrt{\frac{8}{5}}\left(\br_m-\frac{\br_l+\br_k+\br_j+\br_i}{4}
    \right)\\
    \xx_{3p}&=\sqrt{\frac{3}{2}}\left(\br_l-\frac{\br_k+\br_j+\br_i}{3}
    \right)\\
    \xx_{4p}&=\sqrt{\frac{4}{3}}\left(\br_k-\frac{\br_j+\br_i}{2}
    \right)\\
    \xx_{5p}&=\br_j-\br_i\,,\label{eq:jacvec}
  \end{aligned}
  \end{equation}
  where $(i,j,k,l,m,n)$ indicates a generic permutation $p$  of the particles.
  By definition $p=1$ is chosen to correspond to $(1,2,3,4,5,6)$.
  In the following, the ``standard'' set of Jacobi vectors will be
  defined to correspond to this particular definition.

  For a given choice of the Jacobi vectors, the hyperspherical
  coordinates are given by the hyperradius $\rho$, which is independent on the
  permutation $p$ of the particles  and is defined as
  \begin{equation}
    \rho=\sqrt{\sum_{i=1,N}\xi_{ip}^2}\,,\label{eq:rho}
  \end{equation}
  and by a set of variables, which in the Zernike and Brinkman
  representation~\cite{Zernike1935,Fabre1983}, are the polar angles
  $\hxx_{ip}=(\theta_{ip},\phi_{ip})$ of each Jacobi vector and the
  four additional ``hyperspherical'' angles
  $\ph_{jp}$, with $j=2,\dots,5$, defined as
  \begin{equation}
    \cos \ph_{jp}=\frac{\xi_{jp}}{\sqrt{\xi_{1p}^2+\dots+\xi_{jp}^2}}\,,
    \label{eq:phiang}
  \end{equation}
  where $\xi_{jp}$ is the modulus of the Jacobi vector $\xx_{jp}$.
  The set of variables $\hxx_{1p},\dots,\hxx_{5p},\ph_{2p},\dots,\ph_{5p}$
  is denoted hereafter as $\Omega_p$.
  The expression of the generic HH function is
  \begin{equation}\label{eq:hh6}
  \begin{aligned}
    {\cal Y}^{KLM}_{\mu}(\Omega_p)&=\big[((( Y_{\ell_1}(\hat \xi_{1p})
      Y_{\ell_2}(\hat \xi_{2p}))_{L_2} Y_{\ell_3}(\hat \xi_{3p}))_{L_3}
     \\
    &\times  Y_{\ell_4}(\hat \xi_{4p}))_{L_4} Y_{\ell_5}(\hat \xi_{5p})
    \big]_{LM}\\
    &\times{\cal P}^{\ell_1,\ell_2,\ell_3,\ell_4,\ell_5}
         _{n_2,n_3,n_4,n_5}(\ph_{2p},\ph_{3p},\ph_{4p},\ph_{5p})\,,
  \end{aligned}
  \end{equation}
  where
  \begin{equation}\label{eq:hh6b}
  \begin{aligned}
    &{\cal P}^{\ell_1,\ell_2,\ell_3.\ell_4,\ell_5}
    _{n_2,n_3,n_4,n_5}(\ph_{2p},\ph_{3p},\ph_{4p},\ph_{5p})\\
    &={\cal N}_{n_2}^{\ell_2,\nu_2}(\cos\ph_{2p})^{\ell_{2}}
    (\sin\ph_{2p})^{\ell_1}P_{n_2}^{\ell_1+1/2,\ell_2+1/2}(\cos 2\ph_{2p})
       \\
       &\times{\cal N}_{n_3}^{\ell_3,\nu_3}(\cos\ph_{3p})^{\ell_3}
    (\sin\ph_{3p})^{K_2}P_{n_3}^{\nu_2,\ell_3+1/2}(\cos 2\ph_{3p})\\
       &\times{\cal N}_{n_4}^{\ell_4,\nu_4}(\cos\ph_{4p})^{\ell_4}
    (\sin\ph_{4p})^{K_3}P_{n_4}^{\nu_3,\ell_4+1/2}(\cos 2\ph_{4p})\\
       &\times{\cal N}_{n_5}^{\ell_5,\nu_5}(\cos\ph_{5p})^{\ell_5}
    (\sin\ph_{5p})^{K_4}P_{n_5}^{\nu_4,\ell_5+1/2}(\cos 2\ph_{5p})\ ,
  \end{aligned}
  \end{equation}
  and $P^{a,b}_n$ are Jacobi polynomials.
  The coefficients ${\cal N}_{n_j}^{\ell_j,\nu_j}$ are normalization factors
  given explicitly by
  \begin{equation}\label{eq:nhh}
    {\cal N}^{\ell_j,\nu_j}_{n_j}=
    \biggl[{2\nu_j\Gamma (\nu_j-n_j)n_j!\over
        \Gamma (\nu_j-n_j-\ell_j-{1\over 2})
        \Gamma (n_j+\ell_j+{3\over 2})}\biggr]^{1/2}\,,
  \end{equation}
  where we have defined
  \begin{equation}
    K_j=\ell_j+2n_j+K_{j-1}\,,\qquad
    \nu_j=K_j+\frac{3}{2}j-1\,,
  \end{equation}
  with $K_1=\ell_1$ and $K_5=K$. The integer index $\mu$ labels the
  choice of all the hyperangular quantum numbers, namely
  \begin{equation}\label{eq:qn3}
    \mu\equiv\{\ell_1,\ell_2,\ell_3,\ell_4,\ell_5,L_2,L_3,
    L_4,n_2,n_3,n_4,n_5\}\ .
  \end{equation}

  The kinetic energy operator for the six particles, without the center of
  mass (c.m.) motion, can be rewritten in term of the variables $\{\rho,\Omega_p\}$
  as
  \begin{equation}
    \sum_{j=1,5}\nabla^2_j=\left[\frac{\partial^2}{\partial \rho^2}+\frac{14}{\rho}
      \frac{\partial}{\partial\rho}+\frac{\Lambda^2(\Omega_p)}{\rho^2}\right]\,,
  \end{equation}
  where $\Lambda^2(\Omega_p)$ is called grandangular momentum operator and depends
  only on the hyperangular coordinates. 
  The HH functions are the eigenfunctions of this operator, namely
  \begin{equation}\label{eq:Kaut}
    \left[\Lambda^2(\Omega_p)+K(K+13)\right]
    {\cal Y}^{KLM}_{\mu}(\Omega_p)=0\,,
  \end{equation}
  where with $K$ we indicate the eigenvalue of this operator, which is
  usually called grandangular quantum number. 
  
  Our wave function is constructed to have a well defined  total
  angular momentum $J$, $J_z$, parity $\pi$ and isospin $T$
  (in the following, we disregard small admixtures between isospin states).
  Therefore, we define a complete basis of antisymmetrical hyperangular-spin-isospin
  states as follows
  \begin{equation}\label{eq:hhst0}
    \Psi^{KLSTJ\pi}_\al=\sum_{p=1}^{360}\Phi^{KLSTJ\pi}_\al(i,j,k,l,m,n)\ ,
  \end{equation}
  where the sum is over the 360 even permutations $p$ of the particles and
  \begin{equation}
  \begin{aligned}
    &\Phi^{KLSTJ\pi}_\al(i,j,k,l,m,n)=
    \big \{{\cal Y}^{KLM}_{\mu}(\Omega_{p})
    [[[s_is_j]_{S_2} s_k]_{S_3}\\
    &\quad\times[[s_ls_m]_{S_4} s_n]_{S_5}]_S\big\}_{JJ_z}
    [[[t_it_j]_{T_2} t_k]_{T_3}
      [[t_lt_m]_{T_4} t_n]_{T_5}]_{TT_z}\,.
    \label{eq:hhst}
  \end{aligned}
  \end{equation}
  The function ${\cal Y}^{KLM}_{\mu}(\Omega_{p})$ is the HH function defined in
  Eq.~(\ref{eq:hh6}) and $s_i$ $(t_i)$ denotes the spin (isospin) function
  of nucleon $i$. Note that the coupling scheme of 
  the spin (isospin) states does not follow the one of the hyperangular part.
  This particular choice
  simplifies the calculation of the potential matrix elements.
  There are other possible choices for the coupling of the spin (isospin)
  states that can be easily connected to our choice
  through combinations of 6j- and 9j-Wigner coefficients. 
  The total orbital angular momentum $L$ of the HH function
  is coupled to the total spin $S$ to give the total angular momentum $J$, $J_z$,
  while the total isospin is given by $T$, $T_z$. 
  The index $\al$ labels the possible choice of hyperangular, spin and
  isospin quantum numbers, namely
  \begin{equation}\label{eq:alpha}
    \begin{aligned}
    \alpha\equiv\{&\ell_1,\ell_2,\ell_3,\ell_4,\ell_5,
    L_2,L_3,L_4,n_2,n_3,n_4,n_5,\\
    &S_2,S_3,S_4,S_5,T_2,T_3,T_4,T_5\}\,,
    \end{aligned}
  \end{equation}
  compatible with the given values of $K$, $L$, $S$, $T$, $J$, and $\pi$.
  The parity of the state is defined by
  $\pi=(-1)^{\ell_1+\ell_2+\ell_3+\ell_4+\ell_5}$ and we will include in our
  basis only the states such that $\pi$ corresponds to the parity of the nuclear
  state under study.

  The total wave function must be completely antisymmetric under exchange
  of any pair of particles. Therefore, we need to impose antisymmetry on
  each state $\Psi^{KLSTJ\pi}_\al$.
  For example, after the permutation of any pair, the state given in
  Eq.~(\ref{eq:hhst0}) can be rearranged so that
  \begin{equation}\label{eq:hhst1}
    \Psi^{KLSTJ\pi}_\al
    \rightarrow\sum_{p=1}^{360}\Phi^{KLSTJ\pi}_\al(j,i,k,l,m,n)\,.
  \end{equation}
  Therefore, to have antisymmetry it is sufficient to impose
  \begin{equation}\label{eq:ants}
   \Phi^{KLSTJ\pi}_\al(j,i,k,l,m,n)=-\Phi^{KLSTJ\pi}_\al(i,j,k,l,m,n)\,.
  \end{equation}
  Under the exchange of $i\leftrightarrow j$ the Jacobi vector $\xx_{5p}$
  [Eq.~(\ref{eq:jacvec})] changes
  its sign, whereas all the others remain the same.
  Therefore, the HH function ${\cal Y}^{KLM}_{\mu}(\Omega_{p})$ in
  Eq.~(\ref{eq:hh6}) transforms into itself times a factor $(-1)^{\ell_5}$.
  Under the   $i\leftrightarrow j$ exchange,
  the spin-isospin part [see Eq.~(\ref{eq:hhst})]
  transforms into itself times a factor $(-1)^{S_2+T_2}$.
  In conclusion the condition in Eq.~(\ref{eq:ants}) is fulfilled when
  \begin{equation}
    \ell_5+S_2+T_2=\text{odd}\,,
  \end{equation}
  that is the condition we impose on the quantum numbers to obtain only
  antisymmetric states.  

  The number $M_{KLSTJ\pi}$ of antisymmetric functions  $\Psi^{KLSTJ\pi}_\al$
  with fixed $K$, $L$, $S$, $T$, $J$ and $\pi$ 
  in general is very large, due to the high
  number of possible combinations of quantum numbers $\alpha$ that fulfill
  the requirements of antisymmetry and parity. However,
  states constructed in such a way 
  are linearly dependent among each other.  In the expansion of
  the wave function, it is necessary to include the linearly
  independent states only.
  The fundamental ingredient
  to identify the independent states is the knowledge of the norm matrix elements
  \begin{equation}\label{eq:norm}
    N^{KLSTJ\pi}_{\al,\al'}=\bra \Psi^{KLSTJ\pi}_\al | \Psi^{KLSTJ\pi}_{\al'}\ket_{\Omega}\,,
  \end{equation}
  where $\bra\ket_{\Omega}$ denotes the spin and isospin trace and the integration
  over the hyperspherical variables.
  The calculation of the above matrix elements, and also those of the Hamiltonian
  (see below), is
  considerably simplified by using the transformation
  \begin{equation}
    \begin{aligned}
      &\Phi^{KLSTJ\pi}_\alpha(i,j,k,l,m,n)\\
      &\qquad=\sum_{\alpha'}a^{KLSTJ\pi}_{\alpha,\alpha'}(p)
      \Phi^{KLSTJ\pi}_{\alpha'}(1,2,3,4,5,6)\,.\label{eq:tc}
      \end{aligned}
  \end{equation}
  The coefficients $a^{KLSTJ\pi}_{\alpha,\alpha'}(p)$ have been obtained using
  the techniques described in Refs.~\cite{Viviani1998,Dohet2019}, generalized
  to the $A=6$ case.
  Hence, the states $\Psi^{KLSTJ\pi}_\al$ of Eq.~(\ref{eq:hhst0})
  can be written as
  \begin{equation}
    \Psi^{KLSTJ\pi}_\al=\sum_{\al'}A^{KLSTJ\pi}_{\al,\al'}
    \Phi^{KLSTJ\pi}_{\alpha'}(1,2,3,4,5,6)\,,\label{eq:tc1}
  \end{equation}
  where
  \begin{align}
    A^{KLSTJ\pi}_{\alpha,\alpha'}=
    \sum_{p=1}^{360} a^{KLSTJ\pi}_{\alpha,\alpha'}(p)\,.
    \label{eq:tc2}
  \end{align}
  The coefficients $A^{KLSTJ\pi}_{\alpha,\alpha'}$ are called Transformation
  Coefficients (TC) and contain all the properties
  of our basis. Therefore, the knowledge of all of them coincides with the
  knowledge of the entire basis.
  By using them, the matrix element of the norm can be easily obtained
  taking advantage of the orthogonality of the HH basis, namely
  \begin{equation}\label{eq:norm1}
    N^{KLSTJ\pi}_{\al,\al'}=\sum_{\al''}\left(A^{KLSTJ\pi}_{\alpha,\alpha''}\right)^*
    A^{KLSTJ\pi}_{\alpha',\alpha''}\,.
  \end{equation}
  Clearly,
  \begin{equation}\label{eq:norm2}
    \bra \Psi^{KLSTJ\pi}_\al | \Psi^{K'L'S'T'J'\pi'}_{\al'}\ket_{\Omega}=0\,,
  \end{equation}
  if $\{KLSTJ\pi\}\neq\{K'L'S'T'J'\pi'\}$. Once the matrix elements
  $N^{KLSTJ\pi}_{\al,\al'}$ are evaluated, the Gram-Schmidt procedure
  have been  used to find and eliminate the linearly dependent states
  among the various $\Psi^{KLSTJ\pi}_\al$ functions.

  We have found that the number of independent antisymmetric states
  $M'_{KLSTJ\pi}$ is noticeably smaller
  than the corresponding $M_{KLSTJ\pi}$. For example, in Table~\ref{tab:states}
  we report the values of  $M_{KLSTJ\pi}$ and $M'_{KLSTJ\pi}$ for the case
  $J=1$, $T=0$, $L=0$ and $2$ and $\pi=+$, which corresponds
  to the main components of the $\Li$ ground state.
  Observing the table it is possible to notice that
  the values of $M_{KLS01+}$ start to be very
  large already for $K=6$ while those $M'_{KLS01+}$ are much smaller.
  For the case $K=0$ and $LSTJ\pi=0101+$,
  there are no independent state due to the Pauli principle.
    \begin{table*}[t]
      \begin{center}
    \begin{tabular}{r@{$\quad$}r@{$\quad$}r@{$\quad$}r@{$\quad$}
        r@{$\quad$}r@{$\quad$}r@{$\quad$}r@{$\quad$}r@{$\quad$}}
      \hline
      \hline
      \multicolumn{1}{c}{$K$} & \multicolumn{2}{c}{$L=0$ $S=1$}&
      \multicolumn{2}{c}{$L=2$ $S=1$} & \multicolumn{2}{c}{$L=2$ $S=2$}&
      \multicolumn{2}{c}{$L=2$ $S=3$} \\
      &$M_{K0101+}$ &$M'_{K0101+}$ &$M_{K2101+}$ &$M'_{K2101+}$ &$M_{K2201+}$&
      $M'_{K2201+}$ &$M_{K2301+}$ &$M'_{K2301+}$\\
      \hline
      0 &      21&   0 &        &     &        &    &       &     \\
      2 &     306&   1 &     327&    1&     177&   0&     34&   0 \\
      4 &   2,325&   7 &   4,662&   12&   2,562&   4&    504&   1 \\
      6 &  12,480&  34 &  34,065&   90&  18,815&  42&  3,730&   9 \\
      8 &  52,893& 144 & 172,500&  442&  95,500& 227& 19,000&  46 \\
      10 & 187,842& 509 & 684,885& 1535$^*$& 379,635& 804$^*$& 75,670& 145\\
      12 & 580,767&     &2,280,030&        &1,264,730&       &252,360&  \\
      14 &1,605,588&    &         &        &         &       &       & \\
      \hline                                               
      \hline                                               
    \end{tabular}
    \caption{\label{tab:states}
      Number of six-nucleon antisymmetrical hyperspherical-spin-isospin
      state for the case $J=1$, $\pi=+$, $T=0$ and $L=0$ and $2$
      for various cases of the grandangular
      quantum number $K$ and the total spin $S$.
      $M_{KLSTJ\pi}$ is the total number of antisymmetric states while
      $M'_{KLSTJ\pi}$ is the number of independent antisymmetric states.
      With the $^*$ we indicate that the reported value of $M'_{KLSTJ\pi}$
      is underestimated since probably
      there are other independent state that must be included
      to determine the complete basis.}
        \end{center}
    \end{table*}

    The final form of the six-nucleons bound state wave function can be
  written as
  \begin{equation}\label{eq:wf}
    \Psi_6^{J\pi}=\sum_l\sum_{KLST,\alpha}c^{KLST}_{l,\alpha}
    f_l(\rho)\Psi^{KLSTJ\pi}_\al\,,
  \end{equation}
  where the sum is restricted only to the linear independent antisymmetric
  states $\alpha$, and
  $c^{KLST}_{l,\alpha}$ are variational coefficients to be
  determined.
  The hyperradial functions $f_l(\rho)$ are chosen to be
    \begin{equation}
    f_l(\rho)=\gamma^{15/2} \sqrt{\frac{l!}{(l+14)!}}\,\,\, 
    L^{(14)}_l(\gamma\rho)\,\,e^{-\frac{\gamma}{2}\rho} \ ,
    \label{eq:fllag}
  \end{equation}
  where $L^{(14)}_l(\gamma\rho)$ are Laguerre polynomials~\cite{AbramowitzStegun}
  and $\gamma$ is a non-linear variational
  parameter to be optimized in order to have a fast convergence on $l$.
  A typical range for $\gamma$ is $3.5-5.5$ fm$^{-1}$.
  The expansion coefficients $c^{KLST}_{l,\alpha}$ are determined using
  the Rayleigh-Ritz variational principle, obtaining an eigenvalue problem
  then solved by using the procedure of Ref.~\cite{Cullum1981}.

  The most challenging task is the computation of the Hamiltonian
  matrix elements.
  The matrix element of the kinetic energy operator can be obtained analytically
  by exploiting Eq.~(\ref{eq:Kaut}). The matrix element of the  $NN$
  potential results
  \begin{equation}\label{eq:vij2}
    \begin{aligned}
    &V^{KLST,K'L'S'T',J\pi}_{l\al,l'\al'}
    =\\
    &\qquad\quad15\bra f_l\Psi^{KLSTJ\pi}_\al|V_{12}|
    f_{l'}\Psi^{K'L'S'T'J\pi}_{\al'}\ket_{\Omega,\rho}\,,
    \end{aligned}
  \end{equation}
  where $\Psi^{KLSTJ\pi}_\al$ is defined in Eq.~(\ref{eq:hhst0}),
  $f_l$ is defined in Eq.~(\ref{eq:fllag}),  and
  $\bra\ket_{\Omega,\rho}$ denotes the spin and isospin traces and the integration
  over the hyperspherical and hyperradial variables. The factor 15
  takes into account the number of possible pairs in $A=6$ systems.

  In order to compute this matrix element it results convenient to use the
  $jj$-coupling scheme in which the basis state $\alpha$ results
  \begin{equation}
    \Psi_{\al}^{KLSTJ\pi} =
    \sum_\nu B^{KLSTJ\pi}_{\al,\nu} \; \Xi^{KTJ\pi}_{\nu}(1,2,3,4,5,6)
    \label{eq:PSI3jj}\,,
  \end{equation}
  where the new transition coefficients
  $B^{KLSTJ\pi}_{\al,\nu}$ are connected to the coefficients
  $A^{KLSTJ\pi}_{\al,\al'}$ via 6j- and 9j-Wigner coefficients.
  The explicit expression for $\Xi^{KTJ\pi}_{\nu}(1,2,3,4,5,6)$ is
  given by
  \begin{equation}
  \begin{aligned}
    \Xi^{KTJ\pi}_{\nu}&(1,\dots,6)={\cal P}^{\ell_1,\ell_2,\ell_3,\ell_4,\ell_5}
    _{n_2,n_3,n_4,n_5}(\ph_2,\ph_3,\ph_4,\ph_5)\\
    &\times\big\{[(Y_{\ell_5}(\hxx_5)
      (s_1s_2)_{S_2})_{j_1}(Y_{\ell_4}(\hxx_4)s_3)_{j_2}]_{j_{12}}
    [((Y_{\ell_1}(\hxx_1)\\
    &\times Y_{\ell_2}(\hxx_2))_{L_2}
        Y_{\ell_3}(\hxx_3))_{L_3}
      ((s_4s_5)_{S_4}s_6)_{S_5}]_{j_3}
    \big\}_{JJ_z}\\
    &\otimes\big[((t_1t_2)_{T_2} t_3)_{T_3}
      ((t_4t_5)_{T_4} t_6)_{T_5}\big]_{T,T_z}\,.\label{eq:PHIjj}
    \end{aligned}
    \end{equation}
  The index $\nu$ labels all possible choices of the quantum numbers
  \begin{equation}
  \begin{aligned}
    \nu\equiv\{&n_5,\ell_5,S_2,j_1,n_4,\ell_4,j_2,j_{12},\ell_1,\ell_2,\ell_3,
    L_2,L_3,n_2,n_3,\\
    &S_4,S_5,j_3,T_2,T_3,T_4,T_5\}\,,
    \label{eq:nu}
  \end{aligned}
  \end{equation}
  which are compatible with $K,T,J$ and $\pi$.
  Even if in this work we use only two-body forces, this particular
  coupling  scheme results to be very advantageous also
  when the three-nucleon interaction is included.

  In terms of the states expressed in the $jj$-coupling scheme, the
  most generic $NN$ potential matrix element results to be
  \begin{equation}
    \begin{aligned}\label{eq:mxpfinal}
      V^{KLST,K'L'S'T',J\pi}_{l\al,l'\al'}&=15    
      \sum_\nu\sum_{\nu'}B^{KLSTJ\pi}_{\al,\nu}\;B^{K'L'S'T'J\pi}_{\al',\nu'}
      \\
      &\times
      \sum_{T_2^z}C_{T_3,T;T_3',T'}^{T_2,T_5;T_2^z}\;
      v^{K,K',j_1}_{l\nu_y,l'\nu'_y}(T_{2z})\delta_{\nu_x,\nu_x'}\,,
    \end{aligned}
  \end{equation}
  where $\nu_x$ is defined as
  \begin{equation}\label{eq:nux}
    \begin{aligned}
      \nu_x=\{&j_1,n_4,\ell_4,j_2,j_{12},\ell_1,\ell_2,\ell_3,L_2,L_3,\\
      &n_2,n_3,
      S_4,S_5,j_3,T_2,T_4,T_5\}\,,
    \end{aligned}
  \end{equation}
  and
  \begin{equation}
    \nu_y=\{n_5,\ell_5,S_2\}\,.
  \end{equation}
  Note that $\nu\equiv\{\nu_x,\nu_y,T_3\}$.
  Moreover, we have defined the coefficients $C_{T_3,T;T_3',T'}^{T_2,T_5;T_{2z}}$
  which come from the matrix elements of the isospin states,
  as
  \begin{widetext}
  \begin{equation}
    \begin{aligned}
      C_{T_3,T;T_3',T'}^{T_2,T_5;T_{2z}}&=
      \sum_{T_{3z}T'_{3z}}( T_3, T_{3z}, T_5, T_z-T_{3z}|T, T_z)
      ( T'_3, T'_{3z}, T_5, T_z-T'_{3z}|T, T_z)\\
      &\times( T_2, T_{2z}, T_5, T_{3z}-T_{2z}|T_3, T_{3z})
      ( T_2, T_{2z}, T_5, T'_{3z}-T_{2z}|T'_3, T'_{3z})\,,
    \end{aligned}
  \end{equation}
    \end{widetext}
  with $(\cdots|\cdots)$ the Clebch-Gordan coefficients. The
  potential term $v^{K,K',j_1}_{l\nu_y,l'\nu_y'}(T_{2z})$
  is the only part of Eq.~(\ref{eq:mxpfinal})
  which depends explicitly on the locality/non-locality of the potential
  model. In the non-local case it results
    \begin{equation}
    \begin{aligned}\label{eq:mxpint}
      &v^{K,K',j_1}_{l\nu_y,l'\nu_y'}(T_{2z})=
      {\cal N}_{n_5}^{\ell_5,\nu_5}{\cal N}_{n'_5}^{\ell'_5,\nu'_5}
      \int_0^\infty d\rho_5\,\rho_5^{11}\int_0^\infty d\xi_5\, (\xi_5)^2\\
      &\qquad\times\int_0^\infty d\xi'_5\, (\xi'_5)^2
       f_l(\rho)(\cos \ph_5)^{\ell_5}
       (\sin \ph_5)^{K_4}\\
       &\qquad\times
       P_{n_5}^{\nu_4,\ell_5+1/2}(\cos 2\ph_5)
      v^{T_{2z}}_{l_5S_2,l_5'S_2';j_1}(\xi_5,\xi_5')\delta_{S_2,S_2'}\\
      &\qquad\times f_{l'}(\rho')(\cos \ph'_5)^{\ell'_5}
      (\sin \ph'_5)^{K_4}P_{n'_5}^{\nu_4,\ell'_5+1/2}(\cos 2\ph'_5)\,,
    \end{aligned}
    \end{equation}
    where $\rho_5^2=\xi_1^2+\xi_2^2+\xi_3^2+\xi_4^2$,
    $\rho^2=\rho_5^2+\xi_5^2$, $(\rho')^2=\rho_5^2+(\xi_5')^2$,
    $\cos \ph_5=\xi_5/\rho$, $\cos \ph_5'=\xi'_5/\rho'$ and
    $v^{T_{2z}}_{\ell S,\ell'S';j}(\xi_5,\xi_5')$ is the
    non-local two nucleon potential acting between
    two-body states ${}^{2S+1}(\ell)_j$
    and ${}^{2S'+1}(\ell')_j$ with isospin $T_{2z}$.
    The three dimensional integrals are then easily computed
    numerically with high accuracy with standard quadrature techniques.
    The local case can be easily derived by taking into account that
    \begin{equation}
      v^{T_{2z}}_{l_5S_2,l_5'S_2';j_1}(\xi_5,\xi_5')=
      \frac{\delta(\xi_5-\xi'_5)}{\xi_5^2}v^{T_{2z}}_{l_5S_2,l_5'S_2';j_1}(\xi_5)\,.
    \end{equation}
    More details on the algorithm used to compute the $NN$ potential
    matrix elements for the $A=6$ case are given in Appendix~\ref{app:algorithm}.
    
    \section{Choice of the basis}\label{sec:basis}

    The main difficulty of the HH method is the selection of
    a subset of basis states allowing for the best description
    of the nuclear states we are considering. Indeed, although the number of
    independent states is much smaller than the degeneracy of the
    basis, a brute force approach of the method, that is the inclusion
    of all the HH states having $K\leq K_M$ and then increasing $K_M$
    until convergence, would be doomed to fail.
    Moreover, it is very difficult to find all the linearly independent
    states already for values of $K=10$, because of the loss of precision
    in the orthogonalization procedure. For this reason, a good
    selection of a restricted and effective subset of basis state
    is fundamental. Up to now we are limited to values $K_M\leq14$,
    however this permits to reach a reasonable convergence only for
    ``soft'' core potential as the case of the SRG evolved ones.

    It is convenient to separate the HH functions into classes taking
    into account their properties and the fact that the convergence rate
    of each class results rather different.
    The first selection can be done considering the quantity 
    $\ell_{\text{sum}}=\ell_1+\ell_2+\ell_3+\ell_4+\ell_5$. Indeed, the HH states
    with large $\ell_{\text{sum}}$ are less correlated by the $NN$ potential
    because of the centrifugal barrier. The SRG potential have quite weak 
    correlations and so, in our calculation, we can consider
    states with only $\ell_{\text{sum}}\leq4$.
    A second criterion which can be used is to consider the number of particles
    correlated by the HH functions.  The nuclear potential, favors the two-body
    correlations; therefore the HH states which depend only on the
    coordinates of a couple of particles give the main contributions.
    A typical example are HH states with only $n_5$ and $\ell_5$ not zero.
    However, for simplicity, in the following we will
    use only the criterion on $\ell_{\text{sum}}$ for the class definition.
    Moreover, we can divide the $\Li$ ground state in
    $LST$ components. The components allowed by the total
    spin of $\Li$ ground state $J^\pi=1^+$  are given in Table~\ref{tab:LSTgs}.
    Being the $\Li$ ground state an almost pure $T=0$ state, we do
    not consider other isospin states.
    \begin{table}
      \centering
      \begin{tabular}{cccc}
        \hline
        \hline
        $L$ & $S$ & $T$& ${}^{2S+1}L_J$\\
        \hline
        0 & 1 & 0 & ${}^3S_1$\\
        2 & 1 & 0 & ${}^3D_1$\\
        2 & 2 & 0 & ${}^5D_1$\\
        2 & 3 & 0 & ${}^7D_1$\\
        1 & 0 & 0 & ${}^1P_1$\\
        1 & 1 & 0 & ${}^3P_1$\\
        1 & 2 & 0 & ${}^5P_1$\\
        3 & 2 & 0 & ${}^5F_1$\\
        3 & 3 & 0 & ${}^7F_1$\\
        4 & 3 & 0 & ${}^7G_1$\\
        \hline
        \hline
      \end{tabular}
      \caption{\label{tab:LSTgs}$LST$ components 
        of the ground state wave function of $\Li$
        in the spectroscopic notation. The isospin is neglected since
        we impose $T=0$ only.}
    \end{table}

    To study the $\Li$ ground state, we find very convenient to choose the
    classes as described below.
    \begin{itemize}
    \item[a.] Class C1. In this class we include the HH states such that
      $\ell_{\text{sum}}=0$, which belong only to the wave component ${}^3S_1$.
      This class represents the main component of the $\Li$ wave function,
      and in order to obtain a nice convergence we include
      states up to $K_{1M}=14$.
    \item[b.] Class C2. In this class we include the HH states such that
      $\ell_5=2$ and $\sum_{i=1,4}\ell_i=0$.
      This class contains channels belonging
      to all the $D$ waves and  its contribution is fundamental
      to obtain a bound $\Li$.
      For this class we include states up to $K_{2M}=12$.
    \item[c.] Class C3. This class includes all the HH states that
      belong to ${}^3S_1$ component with $\ell_{\text{sum}}=2$.
      This class contains only many-body correlations. Therefore, its 
      impact on the binding energy is less significant.
      For this class we include states up to $K_{3M}=10$.
    \item[d.] Class C4. This class includes all the remaining HH states that
      belong to the $D$ wave with $\ell_{\text{sum}}=2$ and are not included in class C2.
      As class C3, this class contains HH states with $\ell_{\text{sum}}=2$
      and only many-body correlations, therefore
      we expect a similar convergence to class C3.
      For this class we include states up to $K_{4M}=10$.
    \item[e.] Class C5. This class includes all the
      independent HH states which
      belong to the $P$ components up to $K_{5M}=8$
      (states with $\ell_{\text{sum}}=2$ only appears).
      We stop to $K_{5M}=8$ since the contribution of the $P$
      waves to the binding energy is quite tiny.
    \item[f.]Class C6. This class includes all the
      independent HH states which
      belong to the $F$ and $G$  components up to $K_{6M}=8$
      (states with $\ell_{\text{sum}}=4$ only appears).
      We stop to $K_{6M}=8$ since the contribution of the $F$ and $G$
      waves to the binding energy is very tiny.
    \end{itemize}
    
    Moreover, for the
    classes C3 and C4, starting from $K_{iM}\geq10$, we need
    to perform a precision truncation.
    Namely, due to the loss of numerical precision in the orthogonalization
    procedure, states with small orthogonal component generate ``spurious''
    bound states when we diagonalize the Hamiltonian.
    Therefore, we perform a truncation of the basis that
    avoids the generation of these spurious bound states.
    As it will be clear below, the contribution of these classes
    is very small and so the truncation is practically irrelevant
    on the final extrapolation of the binding energy. 
    We want to underline that with this selection of the classes,
    up to $K=8$ the HH basis is complete.
    The convergence is studied as follows. First, only the states
    of class C1 with $K\leq K_{1}$ are included in the expansion and
    the convergence of the binding energy is  studied as the value of $K_1$ is increased
    up to $K_{1M}$.
    Once a satisfactory convergence for the first class is reached, the states
    of the second class with $K\leq K_2$ are added to the expansion
    keeping all the states of the class C1 with $K\leq K_{1M}$. The
    procedure is then repeated for each new class.
    Our complete calculation includes about 7000 HH states.

    \section{Results for the $\Li$ ground state}\label{sec:lithres}
    In this section we report the results obtained for the ground state
    of $\Li$.   In this work we have used
    $\hbar^2/m=41.47$ MeV fm$^2$ for all the potentials. Moreover,
    we use $\gamma=4$ fm$^{-1}$ in the hyperradial functions
    [see Eq.~(\ref{eq:fllag})]. This value has been found optimal
    in order to reach convergence
    to the third decimal digit with a number of Laguerre
    polynomials $l_{max}=16$.
    For all the considered  models,
    when the angular momentum of the pair $j$ is large, the
    $NN$ interactions effect is very small.
    Therefore, all the interactions for $j>6$ are discarded, since their
    effects are negligible as it was already shown in
    Ref.~\cite{Viviani2005} for the $\alpha$ particle.
    
    This section is divided in four parts. The validation of our approach
    is shown in Section~\ref{sec:validation}, comparing our results with the
    ones obtained in Ref.~\cite{Gattobigio2011}.
    In Section~\ref{sec:HHconv}, we discuss
    the convergence of the HH expansion in terms of the various classes
    for the SRG evolved potentials.
    The electromagnetic static properties of $\Li$ ground state are
    considered in Section~\ref{sec:emobs}.
    Finally, the calculation of the $\Li$ asymptotic normalization coefficients
    is presented in Section~\ref{sec:ANC}.

    \subsection{Validation of the results}\label{sec:validation}
    In order to validate our calculation we have performed a benchmark
    with the results presented in Ref.~\cite{Gattobigio2011},
    obtained with the non-symmetrized HH (NSHH) approach.
    We perform the benchmark by using the Volkov potential~\cite{Volkov1965}
    \begin{equation}
      V(r)=V_R{\rm e}^{-r^2/R_1^2}+V_A{\rm e}^{-r^2/R_2^2}\,,
    \end{equation}
    where $V_R=144.86$ MeV, $R_1=0.82$ fm, $V_A=-83.34$ MeV
    and $R_2=1.6$ fm.
    Since the Volkov potential is a central potential, it does not couple the
    different partial wave components of the wave function.
    Therefore, we consider only the $L=0$, $S=1$ and $T=0$ component which
    corresponds to  class C1 and C3. For this study we consider
    states of class C3 up to $K_{3M}=12$.
    In such a way, we are using exactly the same expansion  of
    Ref.~\cite{Gattobigio2011} for $K\leq12$.
    \begin{table}
      \centering
      \begin{tabular}{cccc}
        \hline
        \hline
        $K_{iM}$ & C1 & C1+C3 & Ref.~\cite{Gattobigio2011}\\
        \hline
        2  & $-61.142$ &  $-61.142$ & $-61.142$ \\  
        4  & $-62.015$ &  $-62.015$ & $-62.015$ \\ 
        6  & $-63.377$ &  $-63.377$ & $-63.377$ \\ 
        8  & $-64.415$ &  $-64.437$ & $-64.437$ \\ 
        10 & $-65.310$ &  $-65.354$ & $-65.354$ \\ 
        12 & $-65.823$ &  $-65.884$ & $-65.886$ \\ 
        \hline
        \hline
      \end{tabular}
      \caption{\label{tab:volkov}Binding energy of the bound state of $A=6$
        as function of the grandangular momentum $K$
        obtained with the Volkov potential.
        The first two columns are the results obtained in this work considering
        class C1 and class C1+C3 respectively. In the third column
        we report the results of Ref.~\cite{Gattobigio2011}.}
    \end{table}

    In Table~\ref{tab:volkov}
    we report the binding energy of the bound state of $A=6$ as function
    of the grandangular momentum $K$ for classes C1 and C1+C3.
    As it can be seen from the table, if we use only the HH states
    which belong to class C1 we are not able to reproduce the results
    of Ref.~\cite{Gattobigio2011}, even if only few tens of keV are missing.
    Once we add the HH states belonging
    to class C3 up to $K=10$ we recover the values of Ref.~\cite{Gattobigio2011},
    as we expect, since we are using
    exactly the same basis. As it can be seen, the precision truncation
    performed on the class C3 for $K=10$ is irrelevant.
    This is not the case of  $K=12$,  where a 2 keV difference remains,
    due to the fact we are not including 
    states with $\ell_{\text{sum}}=4$
    because of the truncation precision.
    We want to remark that despite this truncation, only 2 keV are
    missing for $K=12$, well below the precision of the convergence on $K$.

    \subsection{Convergence of the HH expansion}\label{sec:HHconv}
    We study the convergence as explained in Section~\ref{sec:basis}, and the
    results presented are arranged accordingly.
    For example, in Table~\ref{tab:Kconv},
    the binding energy reported in a row with a given set of values $K_1,\dots,K_6$
    has been obtained by including in the expansion all the HH functions
    of class C$i$ with $K\leq K_{i}$, $i=1,\dots,6$. In the following,
    we considered the N3LO500 chiral potential of
    Entem and Machleidt~\cite{Entem2003},
    SRG-evolved with $\Lambda=1.2$, $1.5$, $1.8$ fm$^{-1}$~\cite{Bogner2007}.
    The Coulomb interaction is included as ``bare'' (i.e. not SRG evolved).
    We want to remark that these results are obtained considering only two-body
    forces.

    \begin{table*}
      \centering
      \begin{tabular}{ccccccccc}
        \hline
        \hline
        &  &  &  &  &  &\multicolumn{3}{c}{N3LO500-SRG$\Lambda$} \\
        \cline{7-9}
        $K_1$ & $K_2$ & $K_3$ & $K_4$ & $K_5$ & $K_6$ & $1.2$ fm$^{-1}$
        &$1.5$ fm$^{-1}$ &$1.8$ fm$^{-1}$ \\
        \hline
        $\phz$2  &    &    &    &  &   & 24.779 & 22.315 & 17.946 \\
        $\phz$4  &    &    &    &  &   & 28.606 & 26.779 & 22.656 \\
        $\phz$6  &    &    &    &  &   & 29.714 & 28.395 & 24.646 \\
        $\phz$8  &    &    &    &  &   & 30.030 & 28.937 & 25.425 \\
        10 &    &     &     &   &  & 30.150 & 29.159 & 25.781 \\
        12 &    &     &     &   &  & 30.195 & 29.254 & 25.948 \\
        14 &    &     &     &   &  & 30.213 & 29.295 & 26.031 \\
        &    &     &     &   &  &        &        &           \\
        14 & $\phz$2  &    &    &  &   & 30.263 & 29.362 & 26.108 \\
        14 & $\phz$4  &    &    &  &   & 30.900 & 30.481 & 27.619 \\
        14 & $\phz$6  &    &    &  &   & 31.318 & 31.626 & 29.819 \\
        14 & $\phz$8  &    &    &  &   & 31.413 & 32.006 & 30.827 \\
        14 & 10 &    &    &   & & 31.437 & 32.122 & 31.195 \\
        14 & 12 &    &    &   & & 31.444 & 32.167 & 31.352 \\
        &    &    &    &   &   &     &        &            \\
        14 & 12 & $\phz$6  &    &  &  & 31.445 & 32.168 & 31.354 \\
        14 & 12 & $\phz$8  &    &  &  & 31.477 & 32.210 & 31.396 \\
        14 & 12 & 10 &    &  &   & 31.493 & 32.233 & 31.422 \\
        &    &    &    &  &   &        &        &         \\
        14 & 12 & 10 & $\phz$4  & &  & 31.501 & 32.245 & 31.437 \\
        14 & 12 & 10 & $\phz$6  & &  & 31.550 & 32.329 & 31.548 \\
        14 & 12 & 10 & $\phz$8  & &  & 31.577 & 32.389 & 31.642 \\
        14 & 12 & 10 & 10 & &   & 31.586 & 32.412 & 31.689 \\
        &    &    &    & &  &        &        &         \\
        14 & 12 & 10 & 10 & 2 &  & 31.658 & 32.533 & 31.836 \\
        14 & 12 & 10 & 10 & 4 &  & 31.710 & 32.631 & 31.970 \\
        14 & 12 & 10 & 10 & 6 &  & 31.728 & 32.677 & 32.047 \\
        14 & 12 & 10 & 10 & 8 &  & 31.735 & 32.699 & 32.093 \\
        &    &    &    & &  &        &        &             \\
        14 & 12 & 10 & 10 & 8 & 4  & 31.736 & 32.703 & 32.101 \\
        14 & 12 & 10 & 10 & 8 & 6  & 31.746 & 32.733 & 32.161 \\
        14 & 12 & 10 & 10 & 8 & 8  & 31.750 & 32.751 & 32.209 \\    
        \hline
        \hline
      \end{tabular}
      \caption{\label{tab:Kconv}Convergence of $\Li$ binding energies (MeV)
        corresponding to the inclusion in the wave function of the different
        classes C1--C6, in which the HH basis has been divided.
        The SRG-evolution parameters correspond
        to $\Lambda=1.2$, $1.5$ and $1.8$ fm$^{-1}$. }
    \end{table*}

    We can now analyze the results in Table~\ref{tab:Kconv}.
    We observe that  classes C1 and C2
    are the most important and have the  slowest convergence. Indeed the largest
    values of $K$ must be reached. It is evident that
    increasing the value of the SRG parameter $\Lambda$, the convergence
    becomes slower. This is due to the ``hardness'' of the potential that is
    enhanced when $\Lambda$ is large. Moreover,
    class C2 becomes less and less significant when $\Lambda$ becomes smaller.
    This
    effect is generated by the SRG evolution, which reduces the correlations
    between the $S$- and $D$-waves, when $\Lambda$
    decreases. Even if they are the slowest converging classes, they give
    $98\%$ of the binding energy.
    The contribution of classes C3, C4  is very small for all the
    values of the flow parameters $\Lambda$, and also the convergence
    is much faster. It is very interesting to observe that for both classes the
    contribution to the binding energy depends much less on the value of $\Lambda$
    compared to that of classes C1 and C2. This gives
    an indication that the many-body correlations are not very important, 
    independently on the SRG evolution parameter.
    We find also that classes C5, which corresponds to
    $P$ waves, and C6, which corresponds to $F$ and $G$ waves,
    give very small contributions to the ground state of $\Li$.
    Indeed, in order to obtain the same convergence of the other classes,
    we can stop at $K_{5M}=K_{6M}=8$. 

    Let us comment about the convergence rate as  function
    of the maximum grandangular quantum number $K_{iM}$ of the various classes of
    HH states included in our expansion.
    As shown in various studies~\cite{Zakhareyev1969,Schneider1972,Demin1977,Fabre1983},
    the convergence of the HH functions
    towards the exact binding energy depends primarily on the form of the potential.
    For the chiral potentials, it was observed empirically that
    the convergence rate has an exponential behavior as $K_i$ increases.
    We expect that the same rate of the convergence is obtained also for the
    SRG evolved potentials as
    already observed for example in Ref.~\cite{Jurgenson2011}, even if it was
    obtained within the Harmonic Oscillator basis.

    In order to study the convergence behavior, we indicate with
    $B(K_1,K_2,K_3,K_4,K_5,K_6)$
    the binding energy obtained by including in the expansion all the HH states of
    class C1 with $K\leq K_{1}$, all the HH states of class C2 having
    $K\leq K_{2}$
    and so on. Let us define
    \begin{align}
      \Delta_1(K)&=B(K,K_{2M},0,K_{4M},K_{5M},K_{6M})\nonumber\\
      &\qquad-B(K-2,K_{2M},0,K_{4M},K_{5M},K_{6M})\,,\label{eq:d1}\\
      \Delta_2(K)&=B(K_{1M},K,K_{3M},0,K_{5M},K_{6M})\nonumber\\
      &\qquad-B(K_{1M},K-2,K_{3M},0,K_{5M},K_{6M})\,,\label{eq:d2}\\
      \Delta_3(K)&=B(K_{1M},K_{2M},K,K_{4M},K_{5M},K_{6M})\nonumber\\
      &\qquad
      -B(K_{1M},K_{2M},K-2,K_{4M},K_{5M},K_{6M})\,,\label{eq:d3}\\
      \Delta_4(K)&=B(K_{1M},K_{2M},K_{3M},K,K_{5M},K_{6M})\nonumber\\
      &\qquad
      -B(K_{1M},K_{2M},K_{3M},K-2,K_{5M},K_{6M})\,,\label{eq:d4}\\
      \Delta_5(K)&=B(K_{1M},K_{2M},K_{3M},K_{4M},K,K_{6M})\nonumber\\
      &\qquad
      -B(K_{1M},K_{2M},K_{3M},K_{4M},K-2,K_{6M})\,,\label{eq:d5}\\
      \Delta_6(K)&=B(K_{1M},K_{2M},K_{3M},K_{4M},K_{5M},K)\nonumber\\
      &\qquad
      -B(K_{1M},K_{2M},K_{3M},K_{4M},K_{5M},K-2)\,,\label{eq:d6}
    \end{align}
    where with $K_{iM}$ we indicate that,
    for the class C$i$, we are including all the
    HH states up to the maximum $K$ considered in this work.
    With these definitions, we can compute the ``missing'' binding energy for each class
    due to the truncation of the expansion up to a given $K_{i}$,
    by taking care of the modifications of convergence of a
    class C$i$ due to the inclusion of the other classes.
    Note that for $\Delta_1$($\Delta_2$) we put $K_3=0$($K_4=0$).
    This is because the HH states included in class C3(C4) cannot be added to
    the basis without adding before class C1(C2) due to the orthogonalization
    procedure. For example, we cannot add the HH states of class C3 with $K_3=6$
    without adding before the HH states of class C1 with $K_1=6$.
    Therefore, to have a clear
    convergence pattern for class C1(C2),
    we studied it without adding class C3(C4).
    The changes in the convergence pattern of  class C1(C2) due to the coupling
    with the class C3(C4) are in any case negligible, since class
    C3(C4) gives a very small contribution to the total binding energy.

\begin{figure}
      \centering
      \subfigure[][SRG1.2]{\label{fig:Kconv1}
        \includegraphics[width=\linewidth]{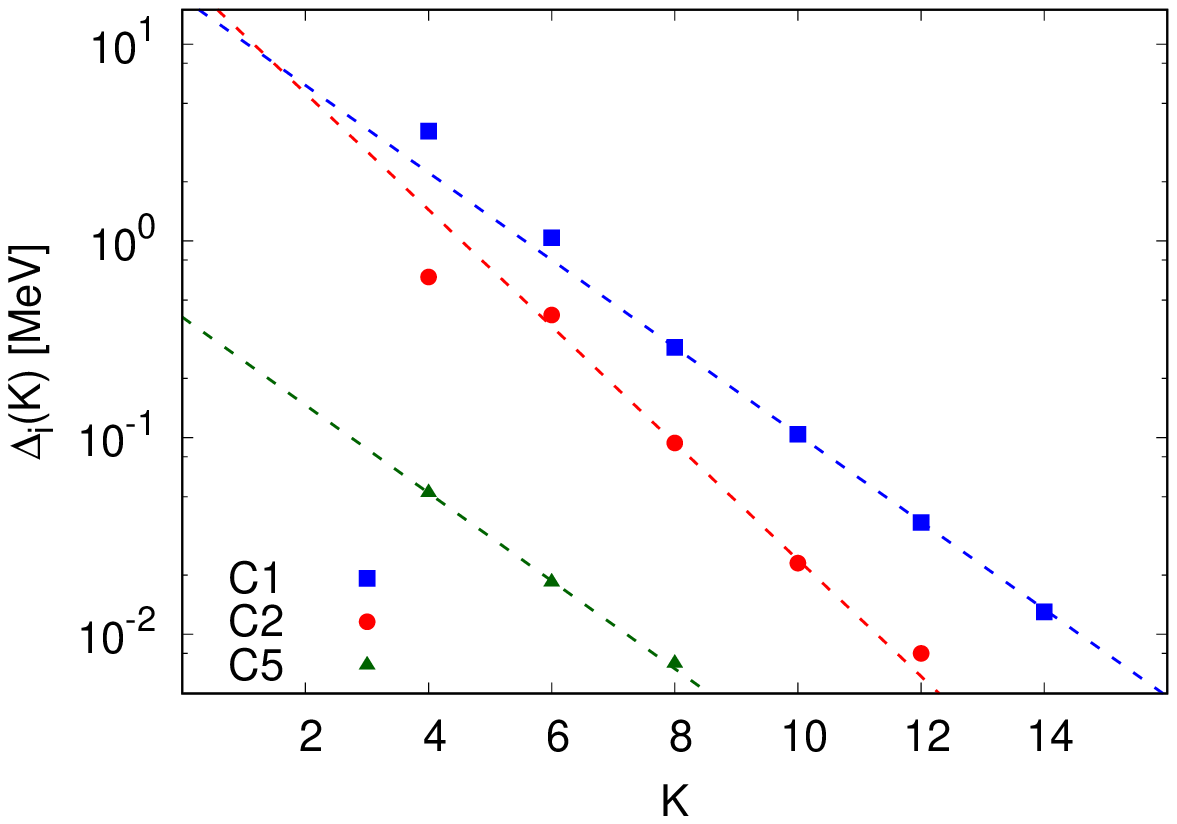}}\\
      \subfigure[][SRG1.5]{\label{fig:Kconv2}
        \includegraphics[width=\linewidth]{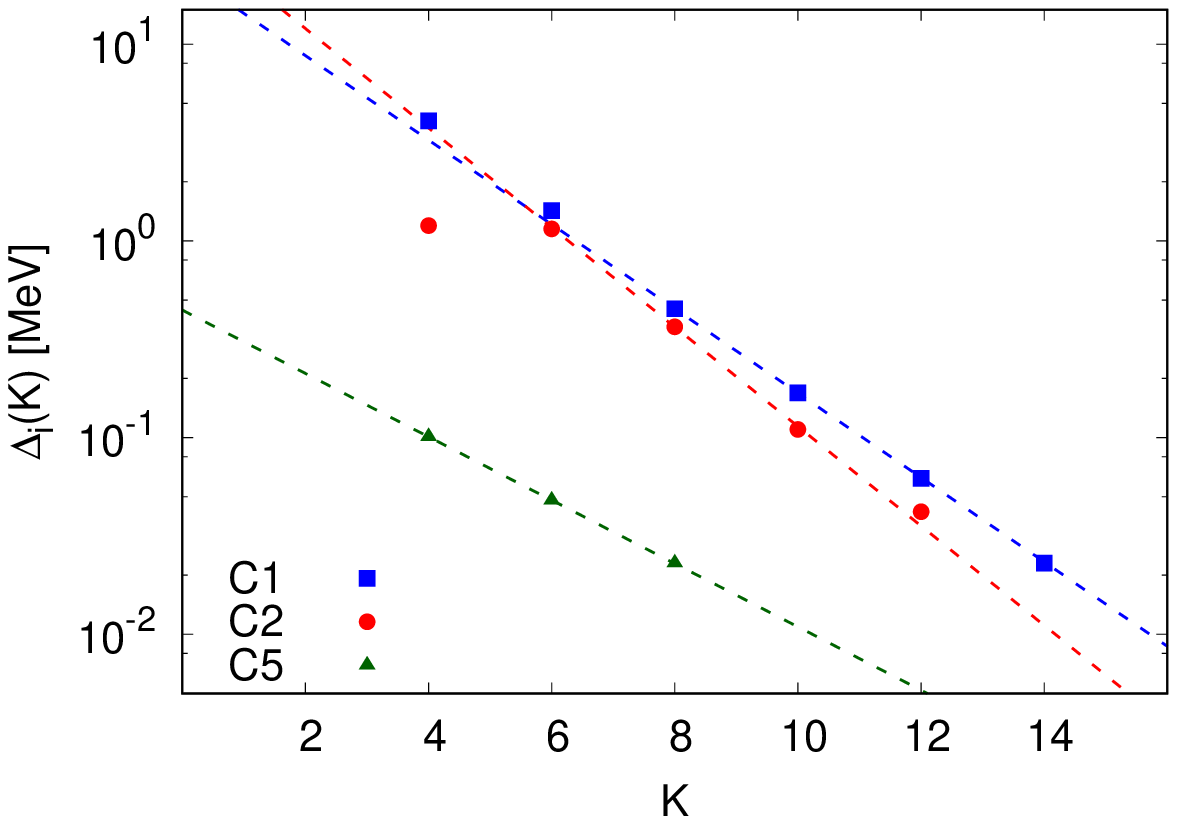}}\\
      \subfigure[][SRG1.8]{\label{fig:Kconv3}
        \includegraphics[width=\linewidth]{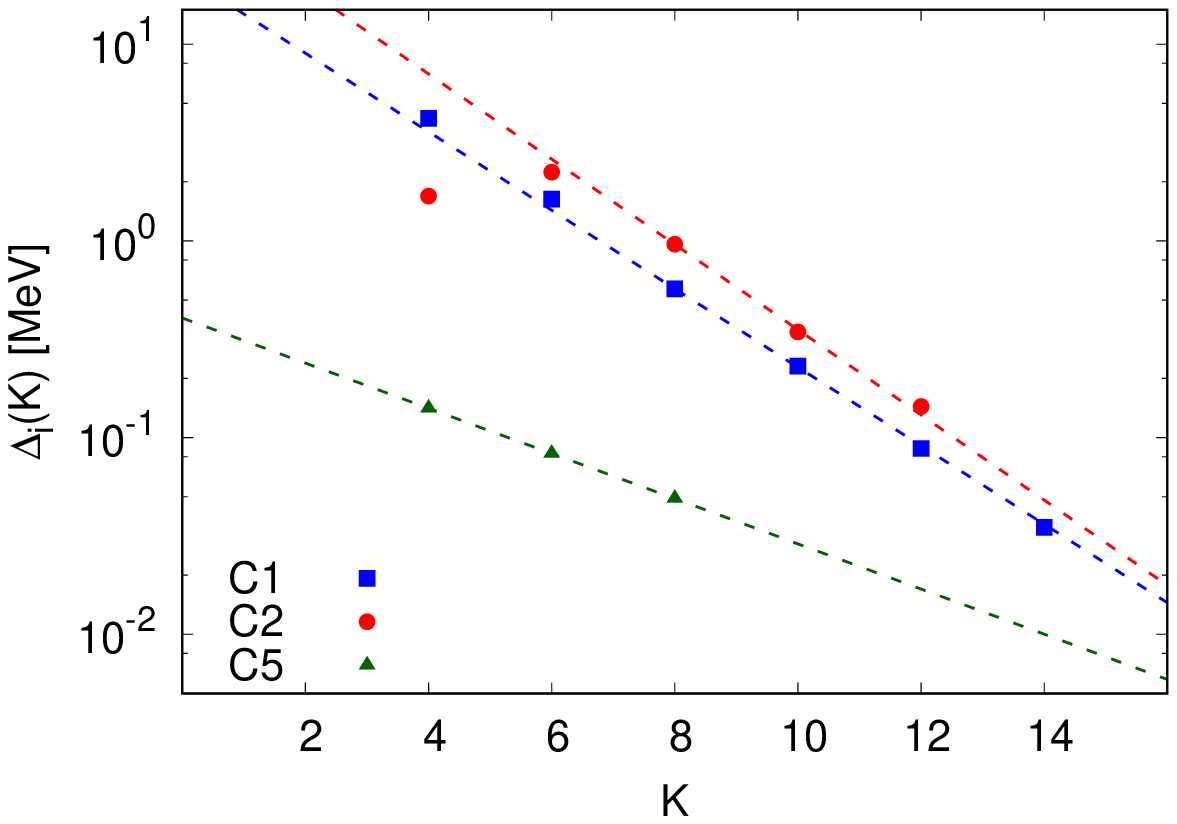}}
      \caption{ Values of $\Delta_i$ [see Eqs.~(\ref{eq:d1})--(\ref{eq:d6})] for the
        classes C1 (squares), C2 (circles), and C5 (triangles) as function
        of the grandangular value K for the three SRG evolved potentials
        $\Lambda=1.2$ fm$^{-1}$ (a), $\Lambda=1.5$ fm$^{-1}$ (b),
        and $\Lambda=1.8$ fm$^{-1}$ (c).
        The dashed lines are the fits obtained using Eq.~(\ref{eq:deltai}).}
      \label{fig:Kconv}
    \end{figure}

    In Figure~\ref{fig:Kconv} we plot the values of
    $\Delta_1$, $\Delta_2$ and $\Delta_5$ for
    the three  SRG evolved potentials considered. By inspecting the figure, we can see
    a clear exponential
    decreasing behavior of the $\Delta_i$ as function of $K$,
    even if  the values of $K$ are rather small. 
    In particular, we can assume for each class that
    \begin{equation}\label{eq:fitb}
      B_i(K)=B_i(\infty)+a_i\,{\rm e}^{-b_iK}\,,
    \end{equation}
    where $B_i(\infty)$ is the asymptotic binding energy of the class
    C$i$ for $K\rightarrow\infty$,
    while $a_i$ and $b_i$ are  parameters which depend on the potential and on
    the class of the HH functions we are studying. In particular, the
    parameter $b_i$ indicates the  convergence rate of the class C$i$.
    From Eq.~(\ref{eq:fitb}) we obtain
    \begin{equation}\label{eq:deltai}
      \Delta_i(K)=a_i{\rm e}^{-b_i K}\left(1-{\rm e}^{2b_i}\right)\,,
    \end{equation}
    which is used  for fitting  $\Delta_i(K)$. The results of the fits
    are the dashed lines in Figures~\ref{fig:Kconv}.
    By observing Figures.~\ref{fig:Kconv1}--\ref{fig:Kconv3}
    it is clear that the convergence rate
    diminishes by increasing the values of $\Lambda$.
    We observe also that for $\Lambda=1.2$ fm$^{-1}$ we have
    $\Delta_1(K)>\Delta_2(K)$, for $\Lambda=1.5$ fm$^{-1}$ we have
    $\Delta_1(K)\approx\Delta_2(K)$ while for $\Lambda=1.8$ fm$^{-1}$ we have
    $\Delta_1(K)<\Delta_2(K)$, which confirms the increasing importance of the
    tensor term of the potential which correlates  $S$- and $D$-waves
    by increasing $\Lambda$.
    Moreover, for all the values of the flow parameter $\Lambda$, we find
    $\Delta_5(K)\ll\Delta_1(K),\Delta_2(K)$, confirming 
    the rapid convergence of the $P$-waves contribution to the binding energy.
    These effects can be seen also by comparing the values of $b_i$ obtained by
    the fits and reported in Table~\ref{tab:missingbe}. 
    For the classes C3, C4 and C6, the calculated values of $\Delta_i(K)$ are not
    enough to perform a fit and so we extract the parameter $b_i$ by using
    only the last two values of $\Delta_i$, namely
    \begin{equation}\label{eq:deltai2}
      \frac{\Delta_i(K_{iM}-2)}{\Delta_i(K_{iM})}={\rm e}^{2b_i}\,.
    \end{equation}
    This formula gives only a rough estimate of the convergence rate.
    The obtained values of $b_i$ are reported in Table~\ref{tab:missingbe}
    as well.
    For all the classes 
    the values of $b_i$ decrease when $\Lambda$ grows which indicates
    a more and more repulsive core of the potential when $\Lambda$ increases.

    Before discussing the calculation of the ``missing'' binding energy, we
    want to underline that Eq.~(\ref{eq:deltai})
    represents the asymptotic behavior of the convergence pattern
    when $K$ is large,
    while we are using value of $\Delta_i$ computed for not so large
    values of $K$.
    For this reason, for the final fit of class C1 and C2
    we used only  $\Delta_i(K)$ with $K\geq8$.
    Indeed, in Figure~\ref{fig:Kconv} it is possible to observe that,
    for $K\leq6$, 
    $\Delta_i$ deviates from the fit. This is usual for the convergence of HH
    states, as already observed in the case of the $\alpha$ particle
    in Ref.~\cite{Viviani2005},
    and is due to the fact that for small values of $K$  the number of states
    are not enough to give a good description of the wave function.
    
    The ``missing'' binding energy due to the truncation of the expansion for each class
    to finite values of $K = K_{iM}$ can be defined as in Ref.~\cite{Viviani2005}
    \begin{equation}\label{eq:dbi}
      \left(\Delta B\right)_i=\sum_{K= K_{iM}+2,K_{iM}+4,\dots}\Delta_i(K)\,,
    \end{equation}
    and, by using Eq.~(\ref{eq:deltai}), we obtain
    \begin{equation}
      \left(\Delta B\right)_i=\Delta_i(K_{iM})\frac{1}{{\rm e}^{2b_i}-1}\,.
    \end{equation}
    The ``total missing'' binding energy is then computed as
    \begin{equation}\label{eq:dbt}
      \left(\Delta B\right)_T=\sum_{i=1,6}\Delta_i(K_{iM})
      \frac{1}{{\rm e}^{2b_i}-1}\,.
    \end{equation}
    In Table~\ref{tab:missingbe} we summarize the ``missing'' binding energy of each class and
    the ``total missing'' binding energy. By inspecting the table we observe
    that the ``total missing'' binding energy is less than $1\%$ of the total binding energy for all the
    SRG evolved potentials.
    This confirms the high accuracy
    of the computed binding energies. 
    As regarding the errors on the ``missing'' binding energy ($\delta(\Delta B)_i$),
    in the case of the class C1, C2 and C5 we propagate the errors
    on $b_i$ evaluated in the fits.
    The estimate of the ``missing'' binding energy
    suffers of the fact that the extrapolation is not really done for large $K$,
    in particular for the class C3, C4 and C6.
    Therefore, for these classes we consider a
    conservative error of  $\delta(\Delta B)_i/(\Delta B)_i=0.5$.
    The error  on the ``total missing'' binding energy is then computed as
    \begin{equation}
      \delta(\Delta B)_T=\sqrt{\sum_{i=1,6}\left(\delta(\Delta B)_i\right)^2}
    \end{equation}
    For all the potentials considered, the relative
    error $\delta(\Delta B)_T/(\Delta B)_T$ is of the order
    of $\sim20\%$.
    
    \begin{table*}
      \centering
      \begin{tabular}{ccccccccccc}
        \hline
        \hline
        &  &\multicolumn{3}{c}{SRG1.2} &\multicolumn{3}{c}{SRG1.5} &
        \multicolumn{3}{c}{SRG1.8} \\
        \hline
        $i$ & $K_{iM}$ &  $\Delta_i(K_{iM})$ &  $b_i$ & $(\Delta B)_i$ &
        $\Delta_i(K_{iM})$ &  $b_i$ & $(\Delta B)_i$ & $\Delta_i(K_{iM})$ &
        $b_i$ & $(\Delta B)_i$\\
        \hline
        1 & 14 & 0.013 & 0.51 & 0.007(0) & 0.023 & 0.49 & 0.014(0)  & 0.035 & 0.46 & 0.023(0) \\ 
        2 & 12 & 0.008 & 0.68 & 0.003(1) & 0.042 & 0.58 & 0.019(0)  & 0.144 & 0.50 & 0.084(11)\\
        3 & 10 & 0.015 & 0.37 & 0.014(7) & 0.022 & 0.32 & 0.024(12) & 0.024 & 0.30 & 0.029(15)\\
        4 & 10 & 0.008 & 0.60 & 0.004(2) & 0.022 & 0.49 & 0.013(6)  & 0.045 & 0.38 & 0.039(20) \\
        5 &  8 & 0.007 & 0.52 & 0.004(0) & 0.023 & 0.37 & 0.021(0)  & 0.049 & 0.26 & 0.070(1) \\
        6 &  8 & 0.004 & 0.44 & 0.003(1) & 0.018 & 0.26 & 0.026(13) & 0.048 & 0.11   & 0.19(9)       \\
        $(\Delta B)_T$ & & &  & 0.034(7) &  & & 0.117(19) & & & 0.43(9)\\
        \hline       
        \hline       
      \end{tabular}  
      \caption{\label{tab:missingbe}Increments of the $\Li$ binding energy
        $\Delta_i(K_{iM})$, computed using Eqs.~(\ref{eq:d1})--(\ref{eq:d6})
        for the various classes
        $i=1,\dots,6$ and the SRG evolved potentials.
        The coefficients $b_i$ are fitted on the $\Delta_i(K)$ for the classes
        $i=1,2,5$ and computed as in
        Eq.~(\ref{eq:deltai2}) for the classes $i=3,4$ and $6$.
        $(\Delta B)_i$ is computed  as in
        Eq.~(\ref{eq:dbi}) and it represents the ``missing''
        binding energy of each class, due to
        the truncation of the expansion up to a given $K_{iM}$. Finally, the
        ``total missing'' binding energy $(\Delta B)_T$ is computed from Eq.~(\ref{eq:dbt}).
        Between the parenthesis we report the errors.
        With $(0)$ we indicate that the
        errors are smaller than the precision of the digits reported in the table.
      }
    \end{table*}
    
    In Table~\ref{tab:comparison} we compare our results with those of
    Ref.~\cite{Jurgenson2011}, obtained using the NCSM.
    As it can be observed by inspecting column one and two,
    the results obtained with
    the same N3LO500-SRG$\Lambda$ potentials in Ref.~\cite{Jurgenson2011},  
    seem to be systematically larger.
    A possible explanation can be found in the fact that in
    Ref.~\cite{Jurgenson2011},  the Coulomb potential is included
    in the SRG evolution.
    By performing the calculations with the Coulomb
    interactions included  in the SRG evolutions
    (indicated with IC in Table~\ref{tab:comparison})
    we gain $\sim30-40$ keV, solving partially
    the discrepancy.  However our results remain still systematically smaller than
    the ones of Ref.~\cite{Jurgenson2011}, even if for $\Lambda=1.2$ fm$^{-1}$
    and $\Lambda=1.8$ fm$^{-1}$ they are compatible within the error bars.
    A possible explanation of the remaining differences  could be
    that we are using a slightly different SRG evolved potential.
    \begin{table}
      \centering
      \begin{tabular}{lccccc}
        \hline
        \hline
        & \multicolumn{2}{c}{NIC (HH)} & \multicolumn{2}{c}{IC (HH)}
        & Ref.~\cite{Jurgenson2011} (NCSM)  \\
        \cline{2-5}
        & $B$ & $B_{ex.}$ & $B$ & $B_{ex.}$ &  $B_{ex.}$\\
        \hline
        SRG1.2  & 31.75   & 31.78(1)   & 31.78 & 31.81(1) &  31.85(5)\\
        SRG1.5  & 32.75   & 32.87(2)   & 32.79 & 32.91(2) &  33.00(5)\\
        SRG1.8  & 32.21   & 32.64(9)   & 32.25 & 32.68(9) &   32.8(1)\\
        \hline
        \hline
      \end{tabular}
      \caption{\label{tab:comparison}Values of the computed ($B$) and extrapolated
        ($B_{ex.}$) $\Li$ binding energy,
        calculated with the HH basis including (IC) and not including (NIC)
        the Coulomb
        interaction in the SRG evolution. Here we report for comparison
        the extrapolated values of Ref.~\cite{Jurgenson2011},
        obtained with the NCSM basis up to $N_{max}=10$. All the results are
        expressed in MeV.}
    \end{table}

    \subsection{Electromagnetic static properties}\label{sec:emobs}
    In order to fully characterize the $\Li$ ground state we compute
    the value of  charge radius, magnetic dipole moment and
    electric quadrupole moment.
    Since the wave function we use is not the
    ``bare'' wave function, we should take care of the SRG
    transformation of the operators in order to be fully consistent.
    However, it has been argued that long-range operators would not be
    affected by it~\cite{Stetcu2005}. Therefore, in this section
    we assume that
    \begin{equation}\label{eq:oos}
      \hat O\approx \hat O(\Lambda)\,,
    \end{equation}
    where $\hat O$ is the ``bare'' operator and $\hat O(\Lambda)$
    the SRG evolved one.
    In any case we will verify this approximation by computing
    the operators for different values of $\Lambda$. 
    Moreover, we discuss the convergence of these observables
    as function of $K$.
    From now on, with $K$  we indicate the fact that for each class
    we include all the HH states with $K_i\leq K$. In the case $K>K_{iM}$
    for a given class C$i$, 
    we include HH states of this class up to $K_{iM}$.

    \subsubsection{Charge radius}
    The mean square (ms) charge radius of a nucleus is given by~\cite{Friar1997}
    \begin{equation}\label{eq:rcdef}
      \bra r_c^2 \ket=\bra r_p^2 \ket+\bra R_p^2 \ket + \frac{N}{Z}\bra R_n^2 \ket
      +\frac{3\hbar^2}{2m_p^2c^2}\,,
    \end{equation}
    where $\bra R_p^2 \ket$ and $\bra R_n^2 \ket$ are the ms charge
    radii of proton and neutron respectively, and the last term
    is the Darwin-Foldy relativistic correction~\cite{Foldy1950}.
    The values used for these
    three contributions are obtained from Ref.~\cite{PDG2018}.
    Moreover, $\bra r_p^2 \ket$ is the ms value of the proton
    point radius operator which for the $\Li$ is defined as
    \begin{equation}
      {\hat r}_p^2=\frac{1}{3}\sum_{i=1}^6(\br_i-\brr_{\text{c.m.}})^2
      \left(\frac{1+\tau_z(i)}{2}\right)\,,
    \end{equation}
    where $\brr_{\text{c.m.}}$ is the c.m. position, $\br_i$ the position
    of the particle $i$, and 3 the number of protons.
    
\begin{figure} 
      \centering
      \includegraphics[width=\linewidth]{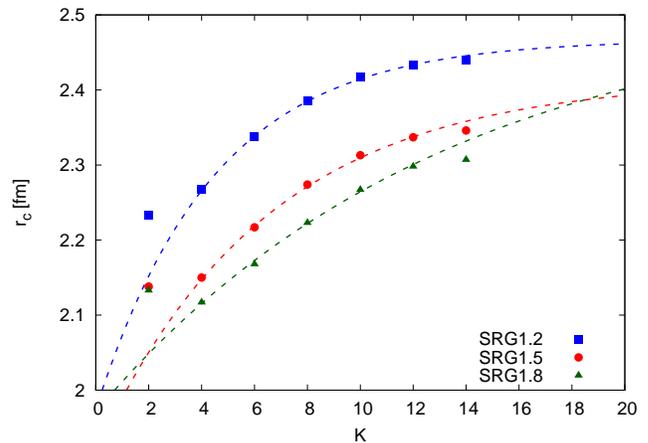}
      \caption{$\Li$ charge radius in fm as function of $K$
      for the potential models
      SRG1.2 (blue), SRG1.5 (red), SRG1.8 (green).
      The exponential fit performed using Eq.~(\ref{eq:fitr}) is also
      shown (dashed lines).}
    \label{fig:radius}
    \end{figure}

    In Figure~\ref{fig:radius} we plot the values of the
    root mean square charge radius $r_c=\sqrt{\bra r_c^2\ket}$
    as function of $K$. From the figure we can observe an exponential behavior
    as $K$ increases. In order to extrapolate
    the  full converged value, we fit our results with
    \begin{equation}\label{eq:fitr}
      r_c(K)=r_c(\infty)+a{\rm e}^{-bK}\,,
    \end{equation}
    where $r_c(\infty)$ is the extrapolated value for $K\rightarrow\infty$.
    The final results are reported in Table~\ref{tab:radius}.
    \begin{table}
      \centering
      \begin{tabular}{rc}
        \hline
        \hline
        & $r_c(\infty)$ [fm]\\
        \hline
        SRG1.2 & 2.47(1) \\
        SRG1.5 & 2.42(2) \\
        SRG1.8 & 2.52(10)\\
        \hline
        Exp. & 2.540(28) \\
        \hline
        \hline
          \end{tabular}
      \caption{\label{tab:radius} Extrapolated values for the $\Li$ charge radii
        obtained using SRG evolved potentials with $\Lambda=1.2$, $1.5$ and $1.8$
        fm$^{-1}$. In the last row
        we report for comparison the experimental value~\cite{Puchalski2013}.}
    \end{table}
    From the fit we have excluded the values of $r_c$ obtained
    for $K=2$ and $K=14$, since they do not follow the exponential behavior,
    as it can be seen in Figure~\ref{fig:radius}. This is due to the fact that
    for $K=2$ there are not enough states to well define the
    structure of $\Li$, and
    among the states with $K=14$ we are not considering channels with  
    $D$ states which are fundamental for describing properly the $\Li$ radius.
    As it can be seen from Figure~\ref{fig:radius}, the
    convergence is quite slow.
    Indeed, the HH basis is a ``compact''
    basis and it is not able to describe perfectly the tail of the wave function
    which has a $\alpha+d$ dominant structure.  
    We will treat this point with more details in Section~\ref{sec:ANC}.
    By comparing the results for the various SRG parameters, it is clear that
    the convergence rate is faster for the smallest values of the
    parameter $\Lambda$,
    since in these cases the correlations between the nucleons are
    reduced, favoring the convergence. The approximation
    of Eq.~(\ref{eq:oos}) for this observable is quite reliable since
    the difference on the extrapolations for the various 
    $\Lambda$ is less than $\sim4\%$.

    \subsubsection{Magnetic dipole moment}

    The magnetic dipole moment operator for the $A=6$ case can be written
    as
    \begin{equation}\label{eq:mudef}
      {\hat \mu_z}=\mu_p{\hat \sigma_z^p}+\mu_n{\hat \sigma_z^n}+{\hat L_z}\,.
    \end{equation}
    where $\mu_p$ and $\mu_n$ are the proton and neutron intrinsic
    magnetic moment taken from Ref.~\cite{PDG2018}, 
    \begin{align}
      {\hat \sigma_z^{p/n}}
      =\sum_{i=1}^6\sigma_z(i)\frac{1\pm\tau_z(i)}{2}\,\label{eq:sz0}
    \end{align}
    is the total spin of protons and neutrons and
    \begin{align}
      {\hat L_z^{p}}&=\sum_{i=1}^6\ell_z(i)\frac{1+\tau_z(i)}{2}\,\label{eq:lz0}
    \end{align}
    is the angular momentum of the protons.
    The convergence of this operator as function of $K$ does not show an
    exponential behavior as in charge radius case but is quite fast since,
    it depends only on the percentage of the various partial waves, which are
    very stable being integral quantities.
    In Table~\ref{tab:mdp2} we report the
    mean values of the magnetic dipole moment
    obtained in the full configuration at $K=14$.
    Since we cannot give a reliable extrapolation to $K\rightarrow\infty$,
    as we have done for the charge radius,
    we consider a  conservative theoretical error defined as 
    \begin{equation}\label{eq:operr}
      \begin{aligned}
        \Delta \mu_z = \max_{K=8,10,12}
        \left\{|\mu_z(K)-\mu_z(14)|\right\}\,,
      \end{aligned}
    \end{equation}
    where $\mu_z(K)$ is the value of the dipole magnetic moment of
    the  $\Li$ wave function computed using $K$ as maximum value for the
    grandangular momentum.
    \begin{table} 
      \centering
      \begin{tabular}{lcc}
        \hline
        \hline
        & $\mu_z(\Li)[\mu_N]$ & $\mu_z(d)[\mu_N]$ \\
        \hline
        SRG1.2  & 0.865(1) & 0.872  \\
        SRG1.5  & 0.858(2) & 0.868  \\
        SRG1.8  & 0.852(2) & 0.865  \\
        \hline
        Exp. &  0.822  & 0.857     \\
        \hline
        \hline
      \end{tabular}
      \caption{\label{tab:mdp2} Values of the
        magnetic dipole moment  $\mu_z$ for $\Li$ 
        evaluated using SRG evolved potential models with $\Lambda=1.2$,
        $1.5$ and $1.8$ fm$^{-1}$.
        We reported also the value of the magnetic dipole moment of deuteron
        $\mu_z(d)$ computed with the same potentials.    
        In the last row the experimental values are listed~\cite{Stone2014}.}
    \end{table}
    As it can be seen by inspecting the table,
    the value of the magnetic dipole moment slightly decreases by
    increasing the value of $\Lambda$. Indeed, when $\Lambda$ increases
    the correlations induced by the nuclear potential
    are stronger, generating a larger amount
    of $D$ component in the wave function, which reduces the value of  $\mu_z$.
    However, the differences between the various $\Lambda$ is $\sim 1\%$
    confirming that Eq.~(\ref{eq:oos}) is a quite good approximation
    for this observable.
    
    If we consider $\Li$ to be formed as a $\alpha+d$ cluster, 
    we can expect that
    \begin{equation}
      \mu_z({\Li})\approx\mu_z(d)\,,
    \end{equation}
    because the $\alpha$-particle has no magnetic dipole moment.
    However, the internal
    structure of $\Li$ plays a fundamental role decreasing
    the value of the magnetic
    dipole moment compared to the deuteron one, as it can be observed
    comparing the experimental
    values (last row of Table~\ref{tab:mdp2}).
    In Table~\ref{tab:mdp2} we compare the results of the magnetic
    dipole moment of
    $\Li$ and $d$ computed with the same SRG potentials.
    In all the cases, 
    the $\Li$ magnetic dipole moment is reduced compared to the $d$ ones,
    showing that the potential models are going in the right direction, even
    if they are not able to reproduce the experimental value.
    Obviously this is partially due to the
    fact we are not considering the evolved operator and also
    that we are not including  three-body forces.
    Moreover, as it was shown in Refs.~\cite{Carlson1998,
      Schiavilla2019} for $\Ht$ and $\Het$,
    the magnetic dipole moment receives important contributions
    from two-body electromagnetic currents.
    Therefore, we can expect that similar corrections are necessary in this case
    to reproduce the experimental value of $\mu_z(\Li)$.

    \subsubsection{Electric quadrupole moment}

    The electric quadrupole moment operator for $A=6$ is defined as
    \begin{equation}\label{eq:qdef}
      {\hat Q}=\sum_{i=1}^6(3z_i^2-r_i^2)\left(\frac{1+\tau_z(i)}{2}\right)\,.
    \end{equation}
    The study of this observable is crucial for understanding the
    goodness of the $\Li$ wave function we computed.
    Indeed, from the experiment, we know that
    the electric quadrupole moment of $\Li$ is very small and negative.
    For this reason,
    it is challenging for all the potential models
    to reproduce this value.

    The convergence of this operator as function of $K$ shows an irregular trend
    due to large cancellations among the contributions coming from different
    sets of HH states.
    Therefore, as for the magnetic dipole moment, we report in
    Table~\ref{tab:quadr} the
    value of this operator obtained for $K=14$ and the errors computed
    as given in Eq.~(\ref{eq:operr}) substituting
    the dipole magnetic moment with the electric quadrupole moment.
    \begin{table}
      \centering
      \begin{tabular}{rc}
        \hline
        \hline
        & $Q$ [$e$ fm$^2$]\\
        \hline
        SRG1.2 & -0.191(7)\\
        SRG1.5 & -0.101(7)\\
        SRG1.8 & -0.055(3)\\
        \hline
        Exp. & -0.0806(6)\\
        \hline
        \hline
      \end{tabular}
      \caption{\label{tab:quadr}
        Values of the $\Li$ electric quadrupole moment obtained for
        SRG evolved potential models with $\Lambda=1.2$, $1.5$ and $1.8$ fm$^{-1}$.
        In the last row
        we report for comparison the experimental value~\cite{Stone2014}.}
    \end{table}
    The values obtained for the SRG potentials
    are quite dependent on the value of the
    parameter $\Lambda$.
    Therefore, for the electric quadrupole moment the approximation
    of Eq.~(\ref{eq:oos}) seems not to be valid.
    However, all the considered SRG evolved potentials are able to reproduce a small and
    negative value for the electric quadrupole moment. In particular, for
    $\Lambda=1.5$  and 1.8 fm$^{-1}$, we obtain values
    quite close to the experimental value of $-0.0806(6)\,e$ fm.

     In order to understand why we have these
     differences between the various SRG evolved potentials,
     we report in Table~\ref{tab:quadr_wave} 
     the partial wave contributions to this observable.
     As it can be seen, we have large differences only in the contribution
     coming from  matrix element between $S$- and $D$-waves.
     In particular the value of this matrix element
     increases when $\Lambda$ increases.
     Therefore, the value of the electric quadrupole moment seems directly connected to the
     strength of the tensor term in the nuclear potential. Indeed, we can expect
     that 
     if the correlations between  $S$- and $D$-waves grow
     (as in the case of ``bare'' chiral potentials), the
     value of the electric quadrupole moment could come positive.
     Therefore, also in this case, the two-body current corrections
     to this observable could be necessary to explain the observed value of
     $\Li$ electric quadrupole moment.
     \begin{table}
       \centering
       \begin{tabular}{lccccc}
         \hline
         \hline
         & $S-D$ & $D-D$ & $P-P$ & $P-D$ & remaining \\
         \hline
         SRG1.2 & $-0.187$ & $-0.023$ & $0.009$ & $0.009$ & $<0.001$ \\
         SRG1.5 & $-0.102$ & $-0.023$ & $0.014$ & $0.010$ & $<0.001$\\
         SRG1.8 & $-0.058$ & $-0.024$ & $0.016$ & $0.010$ & $\phantom{<}0.001$\\
         \hline
         \hline
       \end{tabular}
       \caption{\label{tab:quadr_wave}
         Partial wave contributions to the $\Li$ electric quadrupole moment 
         obtained using SRG evolved potentials with $\Lambda=1.2$, $1.5$ and
         $1.8$ fm$^{-1}$.
         All the values are given in unit of $e$ fm$^2$.}
     \end{table}

     \subsection{Asymptotic Normalization Coefficients}\label{sec:ANC}
     The asymptotic normalization coefficients (ANCs) are properties of
     the bound state wave functions that can be related to experimental
     observables. In particular, in the case of the $\alpha+d$ radiative capture,
     it plays a fundamental role in the determination of the cross section.
     Moreover, the ANCs provide a test of quality of the variational
     wave function in the asymptotic region where the 4+2 clusterization
     is dominant.

     In the asymptotic region, where the $\Li$
     is clustered, the $\Li$ wave function results as
     \begin{align}\label{eq:psiinfty}
       &{\Psi}_{\Li}\rightarrow
       C_0\frac{W_{-\eta,1/2}(2k r)}{r}\Psi_{\alpha+d}^{(0)}
       +C_2\frac{W_{-\eta,5/2}(2k r)}{r}\Psi_{\alpha+d}^{(2)}\,.
     \end{align}
     The function $\Psi^{(L)}_{\alpha+d}$
     is the $\alpha+d$ cluster wave function which is defined as
     \begin{equation}\label{eq:clusterw}
       \Psi^{(L)}_{\alpha+d}=\frac{1}{\sqrt{15}}{\cal A}
       \left[\left(\Psi_\alpha\times\Psi_d\right)_1
         Y_L(\hat r)\right]_1\,,
     \end{equation}
     where the symbol ${\cal A}$ is the antisymmetrization operator,
     $\Psi_\alpha$ and $\Psi_d$ are the wave functions of the $\alpha$-particle
     and the deuteron calculated in the HH variational approach, and
     $\br$ is the distance between the c.m. of the $\alpha$-particle
     and the deuteron. In the previous equation, the spin 0 of the
     $\alpha$-particle
     is combined with spin 1 of the deuteron giving a ``channel'' spin $S=1$.
     The channel spin is then coupled with the angular momentum $L$ to give a
     total angular momentum $J=1$. Because of the even parity of $\Li$, $\alpha$
     and $d$, the $\alpha+d$ cluster can be only in states $L=0,2$.
     In Eq.~(\ref{eq:psiinfty}), $W_{-\eta,L+1/2}(2k r)$ is the Whittaker function
     with $k$ and $\eta$ determined as
     \begin{equation}\label{eq:keta}
       k=\sqrt{\frac{8}{3}\frac{m}{\hbar^2}B_c}\,,\quad
       \eta=\frac{3}{4} \frac{m}{\hbar^2}\frac{e^2}{k}\,.
     \end{equation}
     Here we have used $e^2=1.44$ MeV fm, $\hbar^2/m=41.47$ MeV fm$^2$, and
     $B_c=B_{\Li}-B_{\alpha}-B_d$ with $B_{\Li}$,
     $B_\alpha$
     and $B_d$  the $\Li$, $\alpha$ and $d$ binding energy, respectively.
     Finally, $C_{0/2}$ of Eq.~(\ref{eq:psiinfty}) are the $L=0$ and
     $L=2$ ANCs, respectively.

     In order to compute $C_L$, we have defined the $\alpha+d$ overlap as
     \begin{equation}\label{eq:overlap_df}
       f_{L}(r)=r\frac{\bra \Psi_{\Li}|\Psi^{(L)}_{\al+d}\ket_{r}}
       {\left[\bra\Psi_d|\Psi_d\ket\bra\Psi_\al|\Psi_\al\ket\bra\Psi_{\Li}|\Psi_{\Li}\ket\right]^{1/2} }
       \,,   
     \end{equation}
     where we have defined the proper norms of a generic  wave function of $A$ bodies  as
     \begin{align}
       \bra\Psi_A|\Psi_A\ket&=\int \prod_{i=1}^Ad\br_i\,
       \delta\left(\brr^A_{\text{c.m.}}-\frac{1}{A}\sum_{j=1}^A\br_j
       \right)|\Psi_A|^2=1\,\label{eq:norma},
     \end{align}
     and  $\brr_{\text{c.m.}}^A$ is
     the generic position of the c.m. of the $A$ particles.
     Moreover, in Eq.~(\ref{eq:overlap_df}) we indicate with $\bra\ket_r$
     the fact that we are performing the spin-isospin traces and
     the integration over all the position of the particles
     except the intercluster distance $r$ and the center of mass.
     With such definition
     the overlap is completely independent on the choice of the internal
     variables.
     To perform the calculation it is convenient to introduce
     the proper set of Jacobi coordinates (set ``B'')
     to describe the $\alpha+d$ clusterization,  defined as
     \begin{equation}
       \begin{aligned}
         \xx_{1Bp}&=\br_n-\br_m\\
         \xx_{2Bp}&=\sqrt{\frac{8}{3}}
         \left(\frac{\br_n+\br_m}{2}-\frac{\br_l+\br_k+\br_j+\br_i}{4}
         \right)\\
         \xx_{3Bp}&=\sqrt{\frac{3}{2}}\left(\br_l-\frac{\br_k+\br_j+\br_i}{3}
         \right)\\
         \xx_{4Bp}&=\sqrt{\frac{4}{3}}\left(\br_k-\frac{\br_j+\br_i}{2}
         \right)\\
         \xx_{5Bp}&=\br_j-\br_i\,.\label{eq:jacvecB}
       \end{aligned}
     \end{equation}
     Then, the overlap function reduces to
     \begin{equation}
       \begin{aligned}\label{eq:overlap_df3}
         f_{L}(r)&= r\sqrt{15}
         \left(\frac{\sqrt{6}}{4}\right)^{\frac{3}{2}}
         \int \prod_{i=1}^5d\xx_{iB}\,
         \delta\left(r-\sqrt{\frac{3}{8}}\xi_{2B}\right)\\
         &\times\Psi_{\Li}(\xx_{1B},\xx_{2B},\xx_{3B},\xx_{4B},\xx_{5B},\xx_{6B})^\dag\\
         &\times\left[\left(\Psi_\alpha(\xx_{3B},\xx_{4B},\xx_{5B})\times\Psi_d(\xx_{1B})\right)_1
           Y_L(\hat r)\right]_1\,,
       \end{aligned}
     \end{equation}
     where we have used the antisymmetry of the $\Li$ wave function to eliminate the
     antisymmetrization operator $\cal{A}$, and we have
     multiplied for a factor 15 to take care 
     of the fact that the initial function $\Psi_{\alpha+d}^{(L)}$ contains
     such a number of $4+2$ partitions of the six particles. Finally,
     $\left(\sqrt{6}/{4}\right)^{\frac{3}{2}}$
     is a factor that comes from the normalizations of the wave functions.
     In Eq.~(\ref{eq:overlap_df3})
     we indicate with $\Psi_\alpha(\xx_{3B},\xx_{4B},\xx_{5B})$
     the wave function of the $\alpha$-particle  constructed as the sum over
     the 12 even permutations of the particles $(1,2,3,4)$ and 
     with $\Psi_d(\xx_{1B})$ the wave function of the deuteron constructed with
     the particles $(5,6)$. The $\Li$ wave function is the one of Eq.~(\ref{eq:wf})
     rewritten 
     in terms of the set ``B'' of Jacobi coordinates, by redefining properly
     the TCs.
     The ANCs is then obtained by
     \begin{equation}\label{eq:clr8}
       C_L=\lim_{r\rightarrow\infty}C_L(r)\,,
     \end{equation}
     where
     \begin{equation}\label{eq:clr}
       C_L(r)=\frac{f_L(r)}
       {W_{-\eta,L+1/2}(2k r)}\,.
     \end{equation}

\begin{figure}
       \centering
       \includegraphics[width=\linewidth]{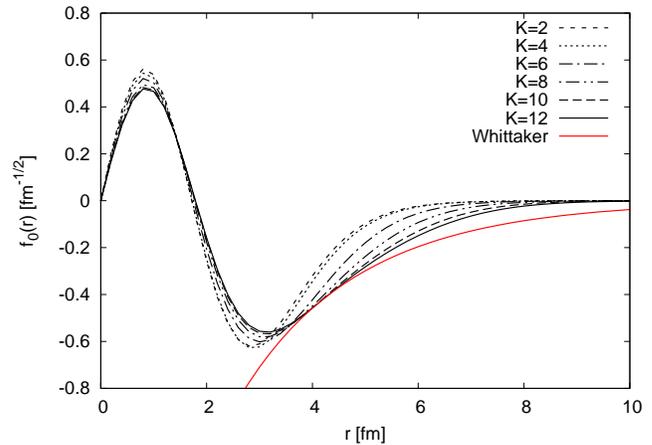}
       \caption{\label{fig:over_conv0}
         $S$-wave component of the overlap function $f_L(r)$ defined in
         Eq.~(\ref{eq:overlap_df3}) for different values of
         $K$ used in the expansion of the $\Li$ wave
         function. The full red line represents the correct asymptotic behavior
         given by the Whittaker function. These results are
         obtained with the SRG1.5 potential.}
     \end{figure}

     The dependence of the overlap
     to the truncation level of the HH expansion of $\Li$
     is studied by varying the maximum value of $K$.
     In Figure~\ref{fig:over_conv0} we plot the $L=0$
     component of the overlap function obtained with the SRG1.5 potential.
     From the figure it is clear that
     the tail of the overlap has not the correct behavior of the Whittaker
     function (full red line). This is
     due to the limited number of HH states used in the expansion
     of the  $\Li$ wave function, which are not enough to reproduce the
     correct asymptotic behavior. However, it is also clear
     that the HH states are slowly constructing the correct asymptotic
     slope when $K$ increases. In reverse, for the
     short-range part ($r<3$ fm) the convergence is fast and completely reached.
     Similar comments
     apply also for all the other potentials and the $L=2$ component.
     The results obtained for the
     $L=0$ and $L=2$ overlaps are qualitative
     consistent with the ones reported in
     Refs.~\cite{Forest1996,Nollet2001,Navratil2004,Navratil2011}.

\begin{figure}
       \centering
       \includegraphics[width=\linewidth]{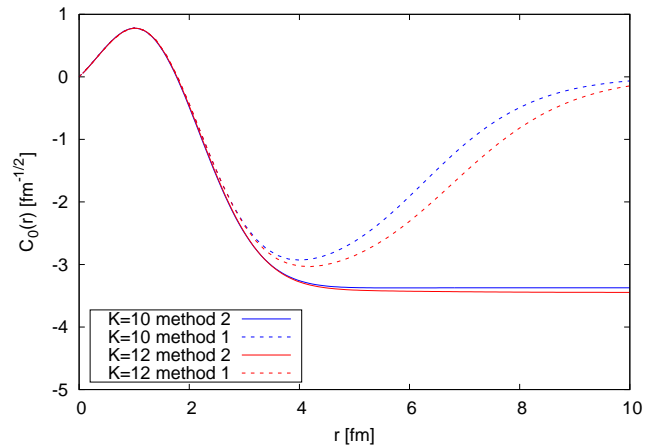}
       \caption{\label{fig:fr0_15}
         Function $C_0(r)$ computed with
         the overlap method (dashed lines) and the equation method (continuous
         lines) for the SRG1.5 potential.
         The calculations are performed with the $\Li$ wave function
         computed with $K=10$ (blue lines) and $K=12$ (red lines). 
         Results of the equation method are obtained using $\kb=8$.}
     \end{figure}

\begin{figure}
       \centering
       \includegraphics[width=\linewidth]{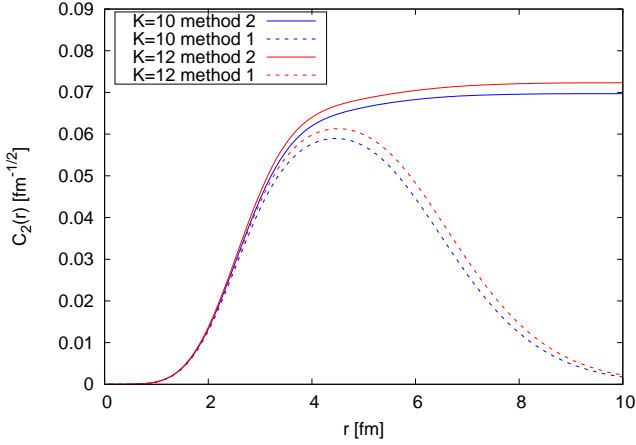}
       \caption{\label{fig:fr2_15} The same as Figure~\ref{fig:fr0_15}, but
         for the function $C_2(r)$.}
     \end{figure}

     By using Eq.~(\ref{eq:clr}), we can then calculate the ANCs.
     In Figure~\ref{fig:fr0_15} and~\ref{fig:fr2_15}
     we illustrate with the dashed lines the 
     ratio $C_0(r)$ and $C_2(r)$ computed in the SRG1.5 case, 
     for $K=10$ (blue) and $K=12$ (red). Both the functions $C_0(r)$
     and $C_2(r)$ shows a sort of
     ``plateau'' around the minimum (maximum in the case $L=2$),
     from which we can have a crude
     estimate of the ANC. We observe the same behavior  for all the
     other potentials considered in this paper.
     In Table~\ref{tab:ANC_comp} with Method I
     we indicate the estimate of the ANCs for both
     $L=0$ and $L=2$ components and the various SRG evolved potentials
     obtained with this approach.
     The numerical differences among the ANCs obtained from the three potentials
     are mostly due to the different values of $k$, present in $\eta$
     [see Eq.~(\ref{eq:keta})],
     entering in the Whittaker function.
     The values of $B_c$, from which the value of $k$ depends,
     are reported as well in
     Table~\ref{tab:ANC_comp}.
     \begin{table*}
       \centering
       \begin{tabular}{llcccc}
         \hline
         \hline
         &Model & $B_c$ [MeV] & $C_0$ [fm$^{-1/2}$] & $C_2$ [fm$^{-1/2}$] &
         $C_2/C_0$ \\
         \hline
         & SRG1.2 & 3.00(1) & $-3.9 $& $0.10$ &$ -0.03 $\\
         Method 1 & SRG1.5 & 2.46(2)&  $-3.0 $& $0.06$ &$ -0.02 $\\
         & SRG1.8 & 2.01(9) & $-2.3 $& $0.03$ &$ -0.01 $\\
         \hline
         & SRG1.2 & 3.00(1) & $-4.19(12)$& 0.116(18)&$ -0.028(5) $\\
         Method 2 & SRG1.5 & 2.46(2) & $-3.44(7) $& 0.072(15)&$ -0.021(5) $\\
         & SRG1.8 & 2.01(9) & $-3.01(7) $& 0.047(10)&$ -0.016(4) $\\
         \hline
         Ref.~\cite{Nollet2001}
         & AV18/UIX   & 1.47     & $-2.26(5) $& -- &$ -0.027(1) $\\
         Ref.~\cite{Hupin2015}
         & SRG1.5(3b) & 1.49     & $-2.695   $& 0.074 & $-0.027$\\
         \hline
         Ref.~\cite{George1999} & Exp.       & 1.4743   & $-2.91(9) $& 0.077(18) & $-0.025(11)$\\
         \hline
         \hline
       \end{tabular}
       \caption{\label{tab:ANC_comp}
         Values of the ANC $C_0$ and $C_2$ in fm$^{-1/2}$ for the
         various SRG evolved potentials with $\Lambda=1.2$, $1.5$ and $1.8$ fm$^{-1}$
         and the two methods used for their calculation.
         We report also the binding energy $B_c$ (MeV)
         used in the calculation of the
         ANC and the ratio $C_2/C_0$.
         For completeness, we shows also the results of the {\it ab-initio} calculation
         of Ref.~\cite{Nollet2001} obtained with AV18/UIX potential, and 
         of Ref.~\cite{Hupin2015} obtained with the SRG1.5 including three-body forces (3b).
         Also the experimental values of Ref.~\cite{George1999} are listed.}
     \end{table*}

     From the overlap it is also possible to compute the spectroscopic
     factor ${\cal S}_L$ defined as
     \begin{equation}
       {\cal S}_L=\int_0^\infty dr\,f_L(r)^2\,,
     \end{equation}
     which can be interpreted as the percentage of $\alpha+d$ clusterization
     in the $\Li$ wave function.
     In Table~\ref{tab:scal} we report the  values of the
     spectroscopic factors obtained for the various potentials.
     Independently on the $\Lambda$ parameter used,
     it results that $\Li$ is clustered in an $\alpha+d$ system for
     more than $80\%$. The differences between the values obtained with
     the SRG evolved potentials
     can be due to the fact that the induced and proper three-body forces
     are not included. Moreover, in the table we compare our results
     with the calculation of Refs.~\cite{Forest1996,Nollet2001,Navratil2004}.
     The values are similar, even if obtained with different potential
     models, and compatible  with the experimental
     estimate of Ref.~\cite{Robertson1981}.
     \begin{table}
       \centering
       \begin{tabular}{llccc}
         \hline
         \hline
         Method & Potential & ${\cal S}_0$ & ${\cal S}_2$ & ${\cal S}_0+{\cal S}_2$\\
         \hline
         \multirow{3}{*}{HH (This work)}& SRG1.2 & 0.909  & 0.008 & 0.917\\
         & SRG1.5 & 0.868  & 0.007 & 0.875\\
         & SRG1.8 & 0.840  & 0.006 & 0.846\\
         \hline
         GFMC (Ref.~\cite{Forest1996})  & AV18/UIX & $0.82$ & $0.021$ & 0.84\\
         GFMC (Ref.~\cite{Nollet2001})  & AV18/UIX & $-$ & $-$ & 0.87(5)\\
         NCSM (Ref.~\cite{Navratil2004})& CD-B2k   & 0.822 & 0.006 & 0.828\\
         Exp. (Ref.~\cite{Robertson1981}) &          & $-$ & $-$ & 0.85(4)\\ 
         \hline
         \hline
       \end{tabular}
       \caption{\label{tab:scal} Values of the spectroscopic factors
         for the various SRG-evolved potentials with $\Lambda=1.2$, 1.5, and
         1.8 fm$^{-1}$. In the last rows 
         we compare our results with the results obtained within the GFMC
         using the AV18/UIX interaction,
         the result obtained within the NCSM using a unitary transformed
         version of the 
         CD-Bonn2000 (CD-B2k) potential~\cite{Machleidt2001}, and
         the experimental one.}
     \end{table}

     Since the procedure adopted so far for extrapolating the ANCs
     results to be somewhat unsatisfactory, due to
     the difficult identification of the ``plateau'',
     we use another procedure,
     based on Ref.~\cite{Timofeyuk1998} and already applied
     in Ref.~\cite{Viviani2005}.
     With this approach, we can extrapolate the ANCs with greater accuracy.
     Assuming that $\Psi_\alpha$ and $\Psi_d$ are ``exact'', it is not difficult
     to show that the overlap function $f_L(r)$ should satisfy the equation 
     \begin{equation}\label{eq:fr}
       \left[-\frac{\hbar^2}{2\mu}\left(\frac{d^2}{dr^2}
         -\frac{L(L+1)}{r^2}\right)
         +\frac{2e^2}{r}+B_c\right]f_L(r)+g_L(r)=0\,,
     \end{equation}
     where $\mu=4/3m$ is the reduced mass of the $\alpha+d$ system, and
     \begin{equation}
       \begin{aligned}\label{eq:GVL}
         g_L(r)&=r\sqrt{15}
         \left(\frac{\sqrt{6}}{4}\right)^{\frac{3}{2}}
         \int \prod_{i=1}^5d\xx_{iB}\,
         \delta\left(r-\sqrt{\frac{8}{3}}\xi_{2B}\right)\\
         &\times\Psi_{\Li}(\xx_{1B},\dots,\xx_{5B})^\dag
         \left(\sum_{i\in\alpha}\sum_{j\in d}V_{ij}-\frac{2e^2}{r}\right)\\
         &\times\left[\left(\Psi_\alpha(\xx_{3B},\xx_{4B},\xx_{5B})
           \times\Psi_d(\xx_{1B})\right)_1
           Y_L(\hat r)\right]_1\,,
       \end{aligned}
     \end{equation}
     is the so called source term with $V_{ij}$ the two-body potential.
     As $r\rightarrow\infty$, the function $g_L(r)\rightarrow
     0$, and the solution of Eq.~(\ref{eq:fr}) coincides with the Whittaker
     function, allowing for the extraction of the ANC using Eq.~(\ref{eq:clr8}).
     
     Since the calculation of $g_L(r)$ in this form is quite involved,
     we want to rewrite it so that we can use the same HH properties
     we used to compute the potential matrix
     element of Eq.~(\ref{eq:vij2}).
     In order to do that, we eliminate the $\delta$-function
     in Eq.~(\ref{eq:GVL}) expanding
      $g_L(r)$ in terms of the Laguerre polynomials, i.e.
     \begin{equation}\label{eq:glvexp}
       g_L(r)=\sum_{n=0}^{N_{\text{max}}}C_n^{L}\overline{f}_n(r)\,,
     \end{equation}
     where
     \begin{equation}
       \overline{f}_n(r)=\gamma_a^{\frac{3}{2}}\sqrt{\frac{n!}{(n+2)!}}
       L_n^{(2)}(\gamma_a r)\mathrm{e}^{-\frac{\gamma_a r}{2}}\,.
     \end{equation}
     The parameter $\gamma_a$ is chosen to optimize the expansion.
     Then, the coefficients $C_n^{L}$ are given by
     \begin{equation}\label{eq:cngl}
       C_n^L=\int dr \,r^2 g_L(r)\overline{f}_n(r)\,,
     \end{equation}
  and, substituting Eq.~(\ref{eq:GVL}) in Eq.~(\ref{eq:cngl}), we obtain
  \begin{equation}
  \begin{aligned}\label{eq:cnglx}
    C_n^L&=\sqrt{15}
    \left(\frac{\sqrt{6}}{4}\right)^{\frac{3}{2}}
    \int \prod_{i=1,5}d\xx_{iB}
    \,\Psi_{\Li}(\xx_{1B},\dots,\xx_{5B})^\dag\\
    &\times\left(\sum_{i\in\alpha}\sum_{j\in d}V_{ij}-\frac{2e^2}{r}\right)
  \big[(\Psi_\alpha(\xx_{3B},\xx_{4B},\xx_{5B})\\
       & \times\Psi_d(\xx_{1B}))_1
      Y_L(\hat r)\big]_1\overline{f}_n(r)
    \Big|_{\br=\sqrt{\frac{3}{8}}{\boldsymbol{\xi}}_{2B}}\,.
  \end{aligned}
  \end{equation}
  If now we define the  4+2 cluster wave function as
  \begin{equation}\label{eq:clustern}
    \Psi_{L,n}=\left[\left(\Psi_\alpha(\xx_{3B},\xx_{4B},\xx_{5B})
      \times\Psi_d(\xx_{2B})\right)_1
      Y_L(\hat r)\right]_1\overline{f}_n(r)\,.
  \end{equation}
  it is possible to expand it in terms of the six-body HH states, namely
  \begin{equation}\label{eq:clustern1}
    \Psi_{L,n}=    \sum_{\lbb=0}^{\lbb_{max}}
    \sum_{\kb=0}^{\kb_{max}}
    \sum_{\lb\,\sbb\,\tb,\bb}c^{\kb\,\lb\,\sbb\,\tb}_{\lbb,\bb}(L,n)
    \sum_{p_\alpha=1}^{12}\Phi^{\kb\,\lb\,\sbb\,\tb}_{\lbb,\bb}(p_\al)
    \,,
  \end{equation}
  where the function $\Phi^{\kb\,\lb\,\sbb\,\tb}_{\lbb,\bb}$
  are the HH functions of Eq.~(\ref{eq:hhst})
  expressed in terms of the set ``B'' of Jacobi coordinates. Note that
  with  $p_\alpha$ we indicate the even permutation of $(i,j,k,l)$ of the
  six particles $(i,j,k,l,5,6)$ which coincide with the 12 permutations of the
  four particles inside the $\alpha$ particle.
  In Eq.~(\ref{eq:clustern1}) 
  the coefficients $c^{\kb\,\lb\,\sbb\,\tb}_{\lbb,\bb}(L,n)$ are obtained
  from
  \begin{equation}\label{eq:cbar}
    c^{\kb\,\lb\,\sbb\,\tb}_{\lbb,\bb}(L,n)
    =\bra \Phi^{\kb\,\lb\,\sbb\,\tb}_{\lbb,\bb}(p=1)|
    \Psi_{L,n}(p=1)\ket_{\Omega_B,\rho}\,,
  \end{equation}
  where with $\bra\ket_{\Omega_B,\rho}$ we indicate the integration
  over all the internal variables and the spin-isospin traces.
  We want to underline that the calculation of these coefficients is very easy
  since it involves only the reference permutation ($p=1$).
  The equivalence in Eq.~(\ref{eq:clustern1}) holds exactly only
  when $\kb_{max}\rightarrow\infty$ and $\lbb_{max}\rightarrow\infty$.
  Obviously this is not the case, but
  we can check the quality of our
  expansion by looking to the
  convergence of the observables when we increase $\kb_{max}$ and $\lbb_{max}$.
  By replacing Eq.~(\ref{eq:clustern1}) in  Eq.~(\ref{eq:cnglx})
  the calculation of the source term reduces to compute a series of
  matrix elements of the potential between HH states, that can be easily done by following
  the procedure of Section~\ref{sec:HHexpansion}.

\begin{figure}
    \centering
    \includegraphics[width=\linewidth]{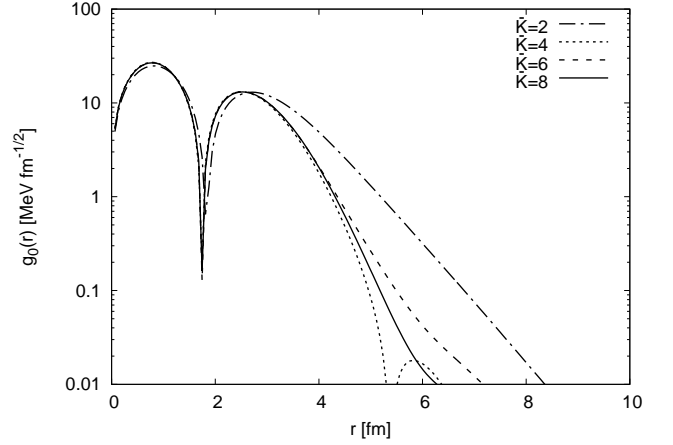}
    \caption{\label{fig:source0}
      The source term $g_0(r)$  for different values of
      $\kb$ used in the projection of the $\alpha+d$ cluster wave functions.
      This figure is obtained by using SRG1.5 potential, and
      the $\Li$ wave function computed with $K=12$.}
  \end{figure}

  For the calculation of the source term we use 
  $N_{\text{max}}=20$ for the expansion  with the Laguerre polynomials
  of the source term [Eq.~(\ref{eq:glvexp})] and $\lbb_{max}=40$
  for the hyperradial part of the cluster function
  [Eq.~(\ref{eq:clustern1})]. Both these values permit to reach full convergence
  in the respective expansions. More interesting is to study the convergence
  as function of $\kb$.
  In Figure~\ref{fig:source0} we plot the source term $g_L(r)$
  for the $L=0$ component
  in the case of the SRG1.5 potential for different values of $\kb$. The
  calculation shown in this plot is performed by considering
  the $\Li$ wave function computed for $K=12$.
  From the figure it is immediately
  clear that we have a nice convergence in $\kb$ for the short-range
  part ($r<3-4$ fm) but not for larger $r$.
  This effect is due to the fact that the Jacobi polynomials
  are not flexible enough
  in reproducing the exponential behavior of the cluster wave functions of
  Eq.~(\ref{eq:clustern}).
  However, by inspecting the figure, it results clear that
  there are problems only in a region where $g_0(r)$ is  a factor 100
  smaller than the peak. A similar convergence behavior in $\kb$
  is found also for the other potentials studied and for $g_2(r)$.

  These considerations on $g_L(r)$ reflects directly on the
  calculated overlap via Eq.~(\ref{eq:fr}).
  Indeed,
  thanks to the fact that the term $g_L(r)$ vanishes for large $r$,
  we obtain the correct asymptotic behavior, namely the Whittaker
  function. 
  In Figures~\ref{fig:fr0_15} and~\ref{fig:fr2_15} we compare
  the functions $C_0(r)$ and $C_2(r)$
  obtained by solving the equation (full lines) with the ones
  obtained from the overlap (dashed lines). In the figures we report the
  results obtained for SRG1.5 where we used
  values of $K=10$ and $12$  for the
  calculation of the $\Li$ wave function and $\kb=8$ for the expansion of the cluster wave function.
  From Figures~\ref{fig:fr0_15} and~\ref{fig:fr2_15}
  it is clear that in the short-range part ($r<2-3$ fm)
  the two approaches are essentially indistinguishable,
  proving the validity of our expansion of the cluster wave function in HH states.
  For larger $r$ the two approaches starts to diverge
  since the functions $C_0(r)$ and $C_2(r)$ computed with the
  direct overlap are not following the correct asymptotic behavior, as already
  discussed. Similar results are obtained also for the other two
  SRG evolved potentials. We observe that the ``harder'' is the
  potential, the smaller is the value of $r$ for which the disagreement
  between the two approaches starts to be important. 
  This is an evidence of the fact
  that also the convergence in $\kb$ depends on the potential.

  From the equation method it is easy now to determine the ANCs with great
  accuracy as function of $K$ and $\kb$ ($C_L(K,\kb)$).
  As function of $\kb$ the ANCs have not a smooth convergence
  and this does 
  not permits us to give a reliable extrapolation for $\kb\rightarrow\infty$.
  Therefore,  we consider as best values the ANCs  obtained at $\kb=8$ to which
  we add  a conservative  error of
  \begin{equation}\label{eq:dclbar0}
    \Delta C^{(0)}_L(K)=0.5\times|C_L(K,\kb=6)-C_L(K,\kb=8)|\,.
  \end{equation}
  As regarding the expansion in $K$, the convergence is smooth and shows
  a clear exponential
  behavior. Therefore we fit the values $C_L(K,8)$ using
  \begin{equation}\label{eq:fitcl}
    C_L(K,8)=C_L(\infty,8)+a{\rm e}^{-bK}\,.
  \end{equation}
  Also for this expansion we 
  consider a conservative error of
  \begin{equation}\label{eq:fiterr}
    \Delta C_L^{(1)}=0.5\times|C_L(12,8)-C_L(\infty,8)|\,.
  \end{equation}
  The total error on the final value $C_L(\infty,8)$ is
  \begin{equation}\label{eq:dclbar}
    \Delta C_L=\sqrt{\left(\Delta C^{(0)}_L(12)\right)^2
      +\left(\Delta C^{(1)}_L\right)^2}\,.
  \end{equation}
  In Table~\ref{tab:ANC_comp} with Method II we indicate our final results
  for $C_L(\infty,8)$. Also in this case
  the numerical differences among the ANCs for the
  three SRG-evolved potentials
  are due to the different binding energy $B_c$
  used to compute the Whittaker functions.
  From  the table it is also evident that the
  results obtained with the equation method are systematically larger than
  the one estimated by the overlap itself. Even if the results obtained with
  the equation are affected by significant errors due to the expansion in
  the HH states   of the cluster wave functions, we consider these as more reliable.
  Indeed, those obtained from the overlap
  suffer from the difficulty of individuating
  unambiguously a well defined plateau and so of unknown systematical errors.
  For completeness in the table we report also the experimental values of Ref.~\cite{George1999}
  and the calculation of Ref.~\cite{Nollet2001}
  obtained with the Argonne $v_{18}$ (AV18) $NN$ interaction~\cite{Wiringa1995}
  combined with the Urbana IX (UIX)
  $NNN$ potential~\cite{Pudliner1997}, and 
  of Ref.~\cite{Hupin2015} obtained with the SRG1.5 potential
  including three-body forces (3b).
  We cannot  really
  perform a comparison, since our results do not contain the contribution
  of the three-body forces, and they are not computed
  at the physical energy $B_c$. However, from a qualitative point of view,
  the results are quite satisfactory,
  since for all the values of $\Lambda$,  we
  are able to reproduce the correct magnitude of the experimental ANCs.

  \section{The role of $\Li$ in the direct dark matter search}\label{sec:dm}
  In recent years experiments devoted to the direct search
  of dark matter are planned to
  use light nuclei, and in particular lithium,
  as probes to search signal of light spin-dependent
  dark matter~\cite{Abdelhameed2019}.
  Usually, in the determination of the sensitivity limit,
  very old shell-model calculations for nuclei are considered.
  However, it is very well known that all shell-models fail
  to describe the $\Li$ structure.
  The aim of this section is to furnish
  reliable calculation of the necessary matrix elements.

  The formula that is usually used to determine the rate of events
  for spin-dependent dark matter
  results to be proportional to $\bra S_{p/n} \ket^2$~\cite{Lewin1996},
  where $\bra S_{p/n}\ket$ is the mean value of the proton (neutron) spin
  operator
  \begin{equation}
    \hat S_{p/n}=\frac{1}{2}\sum_{i=1}^6\sigma(i)\frac{1\pm\tau_z(i)}{2}\,,
  \end{equation}
  on the nuclear wave function.
  In the case of $\Li$, if we consider it
  as a pure state of isospin $T=0$, by exploiting Wigner-Eckart 
  theorem, it is easy to show that
  \begin{equation}
    \bra S_p \ket_{\Li}=\bra S_n \ket_{\Li}\,.
  \end{equation}
  In order to give a very rough estimate of these matrix elements for $\Li$, we
  can consider it as a $\alpha+d$ cluster. If we suppose that $\alpha$
  is a fully spin-0 particle, the only contribution to the spin of $\Li$
  comes from the deuteron, namely
  \begin{equation}\label{eq:spnli}
    \bra S_p \ket_{\Li}=\bra S_n \ket_{\Li}\approx0.5\,.
  \end{equation}

  In our calculation we considered the full $\Li$ structure.
  In particular, by taking care of all the possible partial waves
  which compose the $\Li$ wave function
  (see Table~\ref{tab:LSTgs}), the spin operator results 
  \begin{equation}\label{eq:S_pper}
    \begin{aligned}
    \bra S_{p/n}\ket_{\Li}&=\frac{1}{2}
    P_{{}^3S_1}+\frac{1}{4}P_{{}^1P_1}+\frac{3}{4}P_{{}^3P_1}
    -\frac{1}{4}P_{{}^3D_1}+\frac{1}{4}P_{{}^5D_1}\\&+P_{{}^7D_1}
    -\frac{1}{2}P_{{}^5F_1}+\frac{1}{4}P_{{}^7F_1}-\frac{3}{4}P_{{}^7G_1}\,,
    \end{aligned}
  \end{equation}
  where $P_i$ is the percentage of the $i$-wave in the $\Li$ wave function.
  In Table~\ref{tab:spin} we present the results obtained for the
  proton (neutron) spin with the three different
  SRG evolved potentials used in this work.
  The errors are computed as in Eq.~(\ref{eq:operr}) by substituting the magnetic
  dipole moment with the spin operator.
  \begin{table}
      \centering
      \begin{tabular}{rc}
        \hline
        \hline
        & $\bra S_p \ket(=\bra S_n \ket)$\\
        \hline
        SRG1.2 & 0.479(1)\\
        SRG1.5 & 0.472(2)\\
        SRG1.8 & 0.464(3)\\
        \hline
        \hline
      \end{tabular}
      \caption{\label{tab:spin}
        Mean values of the proton (neutron) spin operator obtained using
        SRG evolved potentials with $\Lambda=1.2$, $1.5$ and $1.8$ fm$^{-1}$.}
  \end{table}
  The differences among the three values and the result of Eq.~(\ref{eq:spnli})
  are directly related to the presence
  of $D$-wave components in $\Li$. Indeed, the larger are the $D$-wave components
  (and in particular the ${}^3D_1$), the smaller is the value of the spin operator. 
  A similar dependence on the percentage of $D$-wave  was
  already found in Ref.~\cite{Korber2017}
  for the nuclear operators that are coupled to spinless dark matter.
  
  \section{Conclusions and perspectives}
  We have studied the solution of the Schr\"odinger equation for the six-nucleon
  ground state using the HH functions.
  The main problem in using the HH approach is the large degeneracy of the
  basis. Therefore, we performed a selection of the HH states that gives
  the most important contributions by following the same procedure
  used in Ref.~\cite{Viviani2005} for the
  $\alpha$-particle ground state. The selection was performed  by dividing
  the HH functions into classes depending on their total angular momentum
  and spin, as well as other quantum numbers.
  For each class we truncate the expansion so as to obtain the required accuracy.

  Many modern $NN$ potentials contain a strong repulsive core which
  makes impossible to reach adequate accuracy with variational approaches
  in $A=6$ systems, because of the huge number of states needed in the wave function expansion.
  Therefore, in this work we limited our study
  to the N3LO500 chiral interaction of Ref.~\cite{Entem2003}
  evolved with the SRG unitary transformation.
  This permits to reach good accuracy with the HH available basis.
  In this first study we considered only two-body forces, even if
  the HH formalism is versatile enough to treat also three-body interactions
  without any additional problem. The inclusion of three-nucleon forces
  is currently in progress.
  
  We have performed the calculation of the binding energy and of
  electromagnetic properties of $\Li$.
  Since we are not considering three-body forces, neither proper, nor induced
  by the SRG evolution, a meaningful comparison of our results with
  the experimental values is still premature.
  Regarding the electromagnetic properties of $\Li$, we have observed
  a strong dependence on the strength of the
  tensor forces. In future, we plan to include also the effect of
  two-body currents, necessary probably
  to explain the small and negative electric quadrupole moment of $\Li$.
  Finally, we have studied the $\alpha+d$ clusterization with the goal of
  determining the asymptotic normalization coefficients. In doing this,
  we performed a projection of the cluster wave function on the HH states. This approach
  can be used also in the study of scattering states, in order to simplify
  the calculation of the potential matrix elements. The calculation
  of the $\alpha+d$ scattering within this approach is also in progress. 

  This work was motivated in order to reach three goals.
  The first one is to show that the HH expansion
  applied to six-nucleon bound problem can reach the same level of precision
  of other approaches, as the NCSM, by using the same potentials. In this sense this work
  represents an  important benchmark for both the HH and the NCSM techniques
  in the  $A=6$ system.
  The second goal is the extension of  this approach to
  work with ``bare'' chiral interactions also in the case of six-nucleon problem.
  All the results reported here were obtained by working on a single core
  with few tens of CPUs. Therefore, we expect possible to consider
  larger basis by using a more massive parallelization. We hope this will allow
  to reach a good accuracy also with a ``bare'' chiral interaction. Moreover, the
  selection of the classes can be improved reducing the number of states
  needed in the expansion.
  The third goal will be the possibility of use this algorithm for systems 
  up to $A=8$. Also in this case a massive parallelization will be fundamental.
  However, also other approaches directly inspired by the NCSM method, as the
  inclusion of clustered component in the wave function,
  can be implemented in the HH formalism in order to improve the convergence.

  \begin{acknowledgments}
    A.G. wish to thanks P. Navr\'atil for useful discussion on NCSM
    results on $\Li$.
    Computations were performed on the MARCONI supercomputer of CINECA in Bologna.
  \end{acknowledgments}
  
  \appendix
  \section{Technical details of the calculation}\label{app:algorithm}
  The biggest computational challenge for applying the HH formalism
  to the $A=6$ system is the calculation and the storage of 
  the potential matrix elements, because of 
  the high number of basis states needed to reach convergence.
  In this appendix we present the main feature of the algorithm
  that we implemented to compute
  the potential matrix elements exploiting the advantages of using the TC.
  The calculations have been performed in parallel machine using a single
  node with 48 Intel Xenon 8160 CPUs @2.10GHz.
  
  Before starting to discuss the algorithm, let us give an idea of
  the dimension of the problem of computing the potential matrix elements
  in this formalism.
  We start from Eq.~(\ref{eq:mxpfinal}).
  The number of
  operations needed to compute the potential matrix elements
  of Eq.~(\ref{eq:mxpfinal}) for given sets $\gamma=\{K,L,S,T\}$ and $\gamma'=\{K',L',S',T'\}$
  and fixed values of Laguerre indices $l$ and $l'$ would be in principle
  \begin{equation}\label{eq:nop}
    N_{op}^{\gamma,\gamma'}\sim N_{\gamma}
    \times N_{\gamma'}\times N_\nu\times N_{\nu'}\,,
  \end{equation}
  where $N_{\gamma}\equiv M'_{KLSTJ\pi}$
  is the total number of independent states
  as defined in Section~\ref{sec:HHexpansion},
  and $N_\nu$ is the number of states entering the expansion given in
  Eq.~(\ref{eq:PSI3jj}),
  which is of the order of $M_{KLSTJ\pi}$.
  For example, if we consider $\gamma=\gamma'=\{12,2,1,0\}$
  which is one of the worst cases, we have
  $N_{\gamma}= N_{\gamma'}\sim 10^3$ and
  $N_\nu=N_{\nu'}\sim 2.3 \times 10^6$ then 
  $N_{op}^{\gamma,\gamma'}\sim 5\times10^{18}$.
  Let us suppose we are in an ideal case in which the time required
  for any of this operation is the typical clock time of a computer,
  $10^{-9}$ s, and that we are able to use in parallel $10^3$ nodes.
  The total time required for doing all these operations is
  \begin{equation}
    T^{\gamma,\gamma'}_{op}\sim 58\,\,{\rm days}\,,
  \end{equation}
  which is a time too long for any practical purpose, especially if we
  need to repeat these operations for all the possible combinations
  of states $\gamma,\gamma'$, of Laguerre polynomials $l,l'$,
  and all the potential models we want to study.
  For this reason we introduce the coefficients $D$, as follows.

  As it can be seen from Eq.~(\ref{eq:mxpint}), the potential integrals
  $v^{K,K',j_1}_{l\nu_y,l'\nu'_y}(T_{2z})$ depends only on the index of the
  Laguerre polynomials and the
  quantum numbers $T_{2z}$, $K$, $K'$, $j_1$, $\nu_y$ and $\nu_y'$, where
  we remember that $\nu_y=\{n_5,\ell_5,S_2\}$.
  Therefore, Eq.~(\ref{eq:mxpfinal}) can be rewritten in a more convenient
  form as
  \begin{align}\label{eq:mxpmodified}
    V^{\gamma,\gamma';J\pi}_{l\al,l'\al'}&=15    
    \sum_{\nu_y,\nu_y'}\sum_{T_3,T_3'}D^{\gamma,\gamma';J\pi}_{\al,\nu_yT_3;
      \al',\nu_y'T_3'}\nonumber\\
    &\times\sum_{T_{2z}}C_{T_3,T;T_3',T'}^{T_2,T_5;T_{2z}}
    v^{K,K',j_1}_{l\nu_y,l'\nu'_y}(T_{2z})\,,
  \end{align}
  where we denote
  $D^{\gamma,\gamma';J\pi}_{\al,\nu_yT_3;\al',\nu_y'T_3'}$ the $D$ coefficient
  and its expression can be easily derived
  comparing Eq.~(\ref{eq:mxpfinal}) with Eq.~(\ref{eq:mxpmodified}).
  Above $\alpha(\alpha')$ defined in Eq.~(\ref{eq:alpha}),
  corresponds to one of the
  $N_\gamma(N_{\gamma'})$ independent states.
  Explicitly the $D$ coefficients are given by
  \begin{align}\label{eq:DDD}
    D^{\gamma,\gamma',J\pi}_{\al,\nu_yT_3;\al',\nu_y'T_3'}=\sum_{\nu_x\nu'_x}
    B^{\gamma J\pi}_{\al,\nu_yT_3\nu_x}
    B^{\gamma'J\pi}_{\al',\nu_y'T_3'\nu'_x}\delta_{\nu_x\nu'_x}\,,
  \end{align}
  where $\nu_x$ are given in Eq.~(\ref{eq:nux}).
  In this way the only part which depends
  on the nuclear interaction in Eq.~(\ref{eq:mxpmodified})
  are the potential integrals $v^{K,K'}_{l\nu_y,l'\nu'_y}(T_{2z})$,
  while the coefficients defined in
  Eq.~(\ref{eq:DDD}) do not.
  Therefore, we can compute and store the $D$ coefficients
  only once for all.
  
  This can be further simplified since for all the possible states
  $\alpha$ and $\alpha'$ with fixed $\gamma$ and $\gamma'$,
  the states $\nu$ and $\nu'$ giving a non vanishing
  contribution are always the same.
  In other words, the determination of the pair of states $\nu,\nu'$ that
  fulfill the condition $\delta_{\nu_x\nu_x'}$,
  which in general requires $N_\nu\times N_{\nu'}$ operations,
  can be performed only once for all the combinations $\alpha,\alpha'$ and
  requires a typical time of $10-20$ minutes for $10\leq K,K'\leq14$
  using a single node with 48 CPUs working in parallel.
  The number of operations which remain to be done in Eq.~(\ref{eq:DDD})
  is then
  equal to the number of states $\nu_x$ ($N_{\nu_x}$). Therefore,
  $N_{op}^{\gamma,\gamma'}$       reduces to
  \begin{equation}
    N_{op}^{\gamma,\gamma'}\sim N_\gamma\times
      N_{\gamma'}\times N_V \times N_{\nu_x}\,,
  \end{equation}
  where $N_V$ is the number of combinations $\nu_y,\nu_y'$ permitted
  by the potential which is typically $<200$.
  The $N^{\gamma,\gamma'}_{op}$ in this case
  is then orders of magnitude smaller than the value
  given in Eq.~(\ref{eq:nop}).
  In a realistic situation, the typical time required for the
  computation of Eq.~(\ref{eq:DDD}), namely 
  to perform the $N_V\times N_{\nu_x}$ operations,
  is $T_D\sim 0.1$ s. Therefore, for $\gamma=\gamma'=\{12,2,1,0\}$
  \begin{equation}
    T_{op}^{\gamma,\gamma'}\sim N_\gamma\times N_{\gamma'}
    \times T_D\sim 1\,\, {\rm day}\,,
  \end{equation}
  using a computer with 48 CPUs on a single node.

  \begin{figure}[h]
    \centering
    \includegraphics[width=\linewidth]{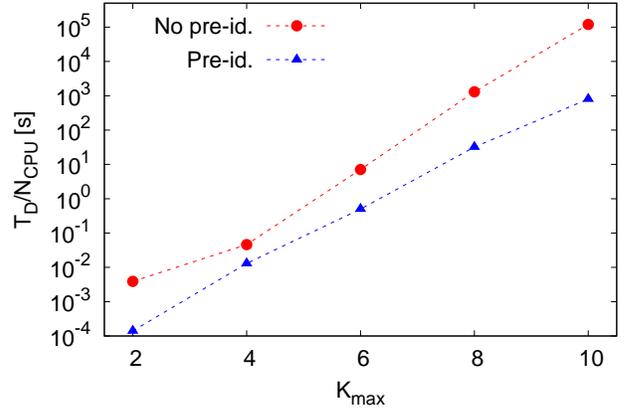}
    \caption{ Total time  needed to compute
      the $D$ coefficients [Eq.~(\ref{eq:ttime})] as function of $K=K'=K_{max}$
      divided by $N_{CPU}$, the number of CPUs used in the computation.
      These calculations are performed for fixed values of the other
      quantum numbers, in particular $L=L'=0$, $S=S'=1$ and $T=T'=0$.
      The red dots are the time spent without pre-identification,
      while the blue triangles are the time spent with pre-identification.
      The dashed lines are added to guide the eyes.
      The calculations were performed on a single node with 48 Intel Xenon
      8160 CPUs @ 2.10 GHz (i.e. $N_{CPU}=48$).}
    \label{fig:ttime}
  \end{figure}
  In Figure~\ref{fig:ttime} we report the total time needed to compute
  the $D$ coefficients when $L=L'=0$, $S=S'=1$ and $T=T'=0$
  up to given $K=K'=K_{max}$, namely
  \begin{equation}\label{eq:ttime}
    T_D=
    \sum_{K=2}^{K_{max}}\sum_{K'=2}^{K_{max}}T_{op}^{K010,K'010}\,.
  \end{equation}
  divided by the number of CPUs ($N_{CPU}$) used in the computation.
  In particular, the blue triangles give $T_D$ by using first
  the pre-identification of the pair of states $\nu,\nu'$ to fulfill the
  $\delta_{\nu_x,\nu_x'}$ condition as discussed before, while the red dots correspond
  to the time spent without pre-identification.
  As it is clear from the figure,
  the computational time increases exponentially by increasing
  the values of $K_{max}$,  since it is
  proportional to the number of independent states which grows
  exponentially as well.
  However, by using the pre-identification, not only $T_D$ results to be well
  reduced, but also as function of $K_{max}$ it has
  a minor slope compared to the case without pre-identification.
  The exponential growth limits the maximum value of $K_{max}$
  we can use at present. However, we expect to have a great improvement
  by using a larger number of CPUs distributing the
  calculation on several nodes.

  As regarding the storage, the total memory required for fixed $\gamma$ and
  $\gamma'$ is given by the number of $D$ coefficients for each
  $\alpha,\alpha'$ combination, namely
  \begin{equation}
    M_{\gamma,\gamma'}[{\rm GB}]\sim 3\times\frac{8\times N_\gamma\times N_{\gamma'}
      \times N_V}{1024^3}\,,
  \end{equation}
  where the factor $3$ is an empirical factor,
  which takes care of the additional information needed
  in the files to save the coefficients $D$.
  For example, when $\gamma=\gamma'=\{12,2,1,0\}$,
  the size of the file is only 2.2 GB.
  The total memory we used to store all the $D$ coefficients used for
  computing the $\Li$ ground state in this work is $\sim 100$ GB.

  Once computed the $D$ coefficients, the time required for the
  calculation of all the potential matrix elements is of the order of a
  couple of hours. Indeed we need only to compute the sum over the
  combinations $\nu_y,\nu_y'$ allowed by the potential ($N_V$), which
  are very few for all the possible $\gamma,\gamma'$ and $l,l'$.
  This process can be even accelerated by computing and storing the matrix element
  of Eq.~(\ref{eq:mxpint}) before combining them with
  the $D$ coefficients.
  
  Typically, in the
  {\it ab-initio} methods, the potential matrix elements are computed
  and stored for each potential model. By  using our approach, we are able
  to save only the $D$ coefficients by eliminating the dependence on
  the potential models.

\bibliography{bibliography}

\end{document}